\documentstyle[psfig,aps]{revtex}

\begin{document}
\title{Dynamics of Coupled Maps with a Conservation Law}
\author{ R. O. Grigoriev and M. C. Cross }
\address{ Condensed Matter Physics 114-36\\
California Institute of Technology, Pasadena CA 91125 }
\date{\today}
\maketitle

\begin{abstract} 
 A particularly simple model belonging to a wide class of coupled maps
which obey a local conservation law is studied. The phase structure of
the system and the types of the phase transitions are determined. It is
argued that the structure of the phase diagram is robust with respect to
mild violations of the conservation law. Critical exponents possibly
determining a new universality class are calculated for a set of
independent order parameters. Numerical evidence is produced suggesting
that the singularity in the density of Lyapunov exponents at $\lambda=0$
is a reflection of the singularity in the density of Fourier modes (a
``Van Hove'' singularity) and disappears if the conservation law is
broken. Applicability of the Lyapunov dimension to the description of
spatiotemporal chaos in a system with a conservation law is discussed. 
 \end{abstract}



\section{Introduction}

Coupled map lattices (CML) with an additive conserved quantity became a
subject of intensive research recently \cite{cross,puri,grinstein}. On the
one hand such CML's are often obtained as phenomenological models
representing the dynamics of a large number of interacting macroscopic
structures. On the other hand they are a natural result of
finite-difference approximations of continuous nonlinear partial
differential equations such as the Kuramoto-Sivashinky equation
\cite{kuramoto}, or a phenomenological Cahn-Hilliard equation \cite{cahn}
describing the nonlinear dynamics of several systems with
conserved-order-parameter. 

Models of this class are expected to represent several typical
non-equilibrium physical phenomena. For instance, surface waves
\cite{tufillaro}, where the average depth of the fluid in the container is
conserved, electrohydrodynamic instabilities in nematics with insulating
plates \cite{nematic}, where the total charge is conserved, disturbances
in the atmosphere and ocean systems, where the total (depth integrated)
heat is conserved, and even some types of hard turbulence \cite{yamada}.
As such they are significant as tools for studying the complex
spatiotemporal behavior of spatially extended nonlinear systems,
especially in the strongly chaotic regime, where the analytical methods
designed for weak nonlinearities become ineffective. 

For coupled map systems with an additive conserved quantity several major
points are still awaiting clarification. First of all, what is the effect
of the conservation law on the structure of the phase diagram and the
character of the phase transitions? What is the connection between the
conservation law and singularities observed in the density of Lyapunov
exponents \cite{cross,bohr,ruelle} and what is the origin and significance
of these singularities? Another important issue is to determine which
parameters best describe the dynamics of an extended system and what their
limitations are. In particular, it is unclear whether the Lyapunov
spectrum provides any exclusive information about the chaotic dynamics
that cannot be obtained by other methods. And finally, we would like to
know whether the reduction of the system dynamics to symbolic form
preserves the main characteristics of the chaotic dynamics and can provide
us with the complete description of the latter. 

The model chosen should be relatively simple yet represent most of the
typical features under consideration. Most important, it should have
nontrivial phase diagram. With this in mind we pick the 1-dimensional
collection of $L$ diffusively coupled chaotic maps: 
 \begin{equation}
 \label{eq_cml}
 u_i^{n+1}=u_i^n+(f(u_{i-1}^n)-2f(u_i^n)+f(u_{i+1}^n))
 \end{equation}
 with periodic boundary conditions imposed. The local map was chosen to be
 \begin{equation}
 \label{eq_map}
 f(x)=ax+bz(1-z),\ z=frac(x).
 \end{equation}

This CML can also be regarded as a finite difference approximation of the
differential equation continuous in both space and time
 \begin{equation}
 \partial_t u(x,t) = \partial_x^2 f(u(x,t)).
 \end{equation}
 A differential equation of this form represents the competition between
two opposing tendencies: generation of chaotic perturbations by the
nonlinear part of $f(x)$ and dissipation of these perturbations by the
diffusive coupling introduced by the second spatial derivative. 

The model clearly possesses the conservation law:
 \begin{equation}
 u = {1\over L} \sum_{i=1}^L u_i^n = const,
 \end{equation}
 so that aside from the two parameters of the local map $a$ and $b$ we
have an additional control parameter -- the additive conserved quantity
$u$, which is defined by the initial condition and is of critical
importance to the behavior of the system. 

We are primarily interested in the dynamics in the ``thermodynamic 
limit'' $L\rightarrow\infty$, although the numerics is obviously 
restricted to finite systems.

The outline of this paper is as follows. In section \ref{sec_phase} we
study the phase diagram of the coupled map model. In section
\ref{sec_symbol} we introduce the symbolic (reduced) description of the
system dynamics. In section \ref{sec_domain} we discuss the quantitative
description of the reduced dynamics. In section \ref{sec_transition} we
study phase transitions in our model and determine their types. In
particular, we study the effect of the conservation law on the type of the
transition and on the values of critical exponents. In section
\ref{sec_lyapunov} we discuss the applicability of the Lyapunov dimension
to the description of the dynamics in the system. In section
\ref{sec_singular} we present numerical data suggesting the reason for the
existence of the singularity in the spectrum of Lyapunov exponents.  In
section \ref{sec_conserv} we demonstrate the effect of violations of the
conservation law on the system dynamics. The paper ends with a summary and
discussion in section \ref{sec_conclusion}. 

\section{Phase Diagram}
\label{sec_phase}

Despite its simple form, this model has a very rich structure.  Numerical
simulations show that depending on the values of the control parameters it
can be strongly or mildly chaotic, show spatiotemporal intermittency
(STI), give rise to pattern formation or simply decay into the spatially
uniform stable state, or (for $L$ -- even) a 2-cycle in both space and
time. Both asymptotically regular (non-chaotic) states can be described by
a single equation: 
 \begin{equation}
 \label{eq_zig}
 u^n_j=u+(-1)^{j+n} A.
 \end{equation}

In order to gain some insight into the phase diagram we analytically
determine the boundaries of the stability regions of the two non-chaotic
asymptotic states of the system. Linear stability analysis of the
spatially uniform state gives Lyapunov exponents (equivalent to the 
growth rates) in the form: 
 \begin{equation}
 \label{eq_lambda}
 \lambda_n=\ln|1-4(a+b-2 b u)\sin^2({\pi n\over L})|.
 \end{equation}
From this one can conclude that the region of linear stability of the 
uniform phase is given by
 \begin{equation}
 \label{eq_l1}
 {a+b\over 2b}-{1\over 4b}<u<{a+b\over 2b}.
 \end{equation}
 Note that at the upper boundary all exponents corresponding to all
Fourier modes $k={2\pi n\over L}$ become real and positive, whereas at the
lower boundary the $k=\pi$ mode goes unstable to a period 2 oscillation. 

Stability analysis of the square of the map (\ref{eq_cml}) gives the
boundaries of the linear stability region of the 2-cycle state written in
a slightly different form: 
 \begin{equation}
 u_i^n=\cases{
 u_{+} &\text{if $i+n$ is even},\cr
 u_{-} &\text{if $i+n$ is odd}.}
 \end{equation} 

The dynamics is invariant under the transformation $u\rightarrow\pm 1$. 
As a consequence it turns out that there are two types of 2-cycles
possible. One (type-I) with
 \begin{equation} 
 [u_{-}]=[u_{+}]-1 
 \end{equation}
 ($[\,\cdot\,]$ denotes the integer part) requires
 \begin{equation}
 u_{\pm}=u(1\pm {2b\over 4bu+1-2a})\label{eq_ampl} 
 \end{equation}
 and has a stability region bounded by the surfaces given by the following
two equations: 
 \begin{equation}
 \label{eq_l2_1}
 (1-2a+2b+4bu_{-})(1-2a-2b+4bu_{+})=-1
 \end{equation}
and
 \begin{equation}
 \label{eq_l2_2}
 u={a\over 2b}.
 \end{equation}
 At the upper boundary (\ref{eq_l2_2}) the $k=\pi$ mode becomes growing,
while at the lower boundary (\ref{eq_l2_1}) a Hopf bifurcation of the
$|k|=\pi/2$ modes occurs. This is quite fortunate, since one can obtain
the analytic expressions for the phase boundaries for a system of
arbitrary size $L$ from the analysis of a system with $L=4$. 

The other (type-II) 2-cycle is such that
 \begin{equation}
 {u_{-}+u_{+}\over 2}=u\ \ \text{and}\ \ [u_{-}]=[u_{+}] \label{eq_oneslot}
 \end{equation}
 and, for the local map $f(x)$ given by eq. (\ref{eq_map}), it can only 
exist at the stability boundary of the uniform state given by
 \begin{equation}
 u={a+b\over 2b}-{1\over 4b},
 \end{equation}
 but can have an arbitrary amplitude $A=(u_{+}-u_{-})/2$, subject only to 
the condition\ (\ref{eq_oneslot}).

Fig.\ \ref{fig_phase} presents two cross sections of the parameter space. 
We will denote the region where the uniform state (1-cycle) is linearly
stable as the phase L1. Similarly, the phase L2 will stand for the linear
stability region of the 2-cycle state. As we are going to see later, the
2-cycle state is not the only possible asymptotic state in this phase, so
it will be useful to introduce the additional subdivision of this phase
into sub-phases for a more detailed analysis. 

The attractors of the phases T1 and T2 are chaotic. The two phases are not
essentially different. One can easily find a continuous trajectory in the
parameter space that would join arbitrary points in phase T1 with those in
phase T2 without intersecting any phase boundary. Although the dynamics of
the system is somewhat different in the two phases for the set of
parameters used, this distinction is introduced mostly for convenience. 

An important property of the system is that despite the large number of
degrees of freedom the type of the attractor (and therefore, the type of
behavior, if one excludes long transients) seems to be uniquely determined
by the values of control parameters $a$, $b$ and $u$ and is independent of
the details of the initial state. If there exists a single attractor, then
the basin of attraction is (almost) all configuration space. In this case 
averages over the attractor can be estimated as the time average from a 
single initial condition, and we can call the system ergodic\footnote{We 
assume that, at least with respect to the numerical computation, there
exists a ``physical measure'' such as discussed by Eckmann and Ruelle
\cite{eckmann} that eliminates any ambiguity in the choice of invariant
measures on the attractor.}.

Indeed, numerical data suggest that the attractor is unique in most of the
parameter space. However it is not always the case: close to the boundary
L1-T1 a frozen pattern may form and the details of a pattern do depend on
the initial conditions. So, on the timescale used in our calculations
(typically of order 1 million iterations) the system did not appear
ergodic. Of course, that does not mean that the system could not become
ergodic on a yet larger timescale. 

Another exception is the regions inside the ordered phases L1 and L2,
where two attractors, chaotic and non-chaotic, can coexist. We will later
discuss this situation in more detail.

One of the objectives of this paper is to study the effect of the
conservation law on the dynamics of the extended chaotic system. We
therefore would like to follow the changes in the parameters
characterizing the dynamics as a function of the conserved quantity $u$.
In further work we will fix the values of the other two parameters at
$a=0.4$, $b=1.3$, which (as seen from the phase diagram, fig. 
\ref{fig_phase}) will allow us to study the regimes of interest (various
chaotic as well as periodic states). 

Looking at the phase diagram, one can expect that this model should
experience at least four bifurcations or phase transitions\footnote{Since
we are interested in diverging correlation lengths near the transitions
between different states in the $L\rightarrow \infty$ limit, and the
consequent possibility of universal exponents, we will use the term
``phase transitions'' rather than ``bifurcations''.} as $u$ is varied in
the interval $0<u<1$. Equations (\ref{eq_l2_1}) and (\ref{eq_l2_2}) give
us $u_a\approx 0.0520$ and $u_b=u_d-0.5\approx 0.1538$ as the boundaries
of the 2-cycle stability region. According to eq. (\ref{eq_l1}) the
uniform state loses its stability at $u_c\approx 0.4615$ and $u_d\approx
0.6538$. 

Phase transitions from ordered to chaotic states are common occurrences in
coupled map lattices \cite{kaneko}. They may be continuous or
discontinuous. For continuous transitions we might expect the transitions
to fall into various universality classes, characterized by scaling
exponents for various diagnostics of the chaos near the transition. In
this case the transitions in CML's may be representative of transitions to
spatiotemporal chaos in more general extended non-equilibrium systems.
Although the qualitative features determining the universality classes are
not understood one expects that symmetries, such as the Ising symmetry
studied by Miller and Huse \cite{miller}, and conservation laws, rather
than the detailed properties of the local maps, will be important. 

In particular, it has been suggested \cite{grassberger_2} that under
certain very general conditions (e.g. in systems with a unique absorbing
state) the transition should fall into the universality class of directed
percolation, although some counter examples to this statement are known.
It is nevertheless interesting to check whether any of the phase
transitions in our model belong to the universality class of directed
percolation, since, as we are going to see below, the absorbing state in
our model is in fact unique for any choice of control parameters. 

\section{Reduced Dynamics}
\label{sec_symbol}

In order to understand the spatial dynamics of the system better and to
see the finer details of the phase diagram we will (following Kaneko \cite
{kaneko_1}) reduce the description of the dynamics to a finite number of
states: in terms of this reduced dynamics each site of the lattice can be
marked either ``laminar'' or ``turbulent'' thus making up a set of laminar
and turbulent domains. Then one would naively expect a laminar domain to
be a region of the lattice with a relatively slow chaotic dynamics (no
large, if any, positive local Lyapunov exponents \cite{kaneko_2}); and a
turbulent domain to be a region where the chaotic dynamics is fast (with
at least a few large positive local Lyapunov exponents). Positive Lyapunov
exponents will inevitably make the turbulent domains spatially irregular,
while laminar domains tend to be spatially regular. 

We will not put the terms ``laminar'' and ``turbulent'' in quotes below,
nevertheless one should clearly understand that these are just a convenient
notation and thus are restricted in meaning.

We need a simple criterion that will determine whether a given site
belongs to a turbulent or laminar domain (we will only consider as laminar
states those that are close to either uniform or 2-cycle configurations).
The simple way to distinguish between (uniform) laminar and turbulent
sites numerically would be to call a site $j$ laminar on time step $n$ if
 \begin{equation}
 |u^n_{j-1}-u^n_j|<\epsilon \text{ and } |u^n_j-u^n_{j+1}|<\epsilon
 \end{equation}
 and turbulent otherwise. The problem with this definition is that any
2-cycle with amplitude 
 \begin{equation}
 A^n_j={|u^n_{j-1}-u^n_j|\over 2} > \epsilon
 \end{equation} 
 would be considered turbulent, which is clearly incorrect. Therefore, the
definition of a laminar domain has to be generalized to include a zig-zag
pattern with slowly varying envelope of arbitrary amplitude: 
 \begin{equation}
 {|A^n_j-A^n_{j+1}|\over |A^n_j+A^n_{j+1}|}<\epsilon.
 \end{equation}
In this particular model we set $\epsilon\sim 0.01$.

Based on this reduction one can distinguish between various types of
dynamics in the system. Pictures representing the spatiotemporal evolution
of the system in symbolic form are so characteristic that one can easily
determine which part of phase diagram the system is in just by looking at
the patterns. We will examine several typical pictures, highlighting the 
most interesting phenomena observed in this model.

The behavior of our CML in the phase L2 is nontrivial. We already know
that inside the phase there is a stable ordered state. Our numerical data
imply however that this ordered 2-cycle state is stable in the nonlinear
sense only for values of u satisfying $u_a<u_p<u<u_n<u_b$ (figure
\ref{fig_hyst}). As we are going to see later, it is quite hard to
determine the critical values $u_p$ and $u_n$ exactly, but they seem to
approach the fixed values $u_p\approx 0.063$ and $u_n\approx 0.082$ in the
thermodynamic limit $L\rightarrow \infty$. 

For $u_p<u<u_n$ the ordered state is reached through a chaotic transient
whose lifetime is usually very small, no matter what initial condition is
chosen, and thus the asymptotic behavior dominates. For relatively small
systems ($L\lesssim 400$) this results in a fast decay of any initial
state into a limit 2-cycle. For larger systems the quiescent asymptotic
state may be completely regular as well, but it may also have a few
localized turbulent defects moving with unit speed through the homogeneous
2-cycle background. Defects moving in opposite directions eventually die
out colliding with each other. Similar properties were observed in a
number of 1- and 2-dimensional models featuring spatiotemporally
intermittent dynamics (see \cite{chat_man_2} and references therein for
example). 

For $u_a<u<u_p$ as well as for $u_n<u<u_b$ the ordered state is only
conditionally stable (stable to small perturbations) and most initial
conditions result in a spatiotemporally chaotic asymptotic state
consisting of a combination of laminar and turbulent domains, with the
laminar domains featuring exactly the same structure as the ordered state:
the type-I 2-cycle. Therefore the phase L2 can be subdivided into three
sub-phases according to whether any turbulent domains are present in the
asymptotic state together with the laminar background whose structure is
the same throughout the phase L2. 

Fig. \ref{fig_hyst} shows schematically the (time averaged) fraction of
the lattice occupied by turbulent domains in the phase L2 as a function of
parameter $u$. In the sub-phase $L2_l$ ($u_p<u<u_n$) the asymptotic state
is laminar. In the sub-phases $L2_p$ ($u_a<u<u_p$) and $L2_n$
($u_n<u<u_b$) the asymptotic state can be either laminar or
spatiotemporally chaotic. It is interesting to note that the
characteristic patterns produced by the turbulent domains are different in
the two sub-phases featuring a persistent spatiotemporally intermittent
state. As a result one might expect to see two quite different phase
transitions at $u=u_p$ and $u=u_n$. 

A pattern typical for the sub-phase $L2_p$ is presented on fig.\
\ref{fig_pattern}(a). This state is very similar to some of the
spatiotemporally intermittent states of a (1-dimensional) model studied by
Chate and Manneville \cite{chat_man}. The major difference is that in our
model turbulent domains typically have a larger size. Since the absorbing
(laminar) state is an ordered one and is unique, we might expect a phase
transition at $u_p$ to belong to the universality class of directed
percolation \cite{grassberger}. 

In fact, this kind of STI state is not specific to discrete extended
systems. A very similar state can be observed in some continuous models as
well, for instance \cite{manneville}, in a damped Kuramoto-Sivashinsky
equation: 
 \begin{equation}
 \partial_t u(x,t)=-\eta u-\partial_x^2 u-\partial_x^4 u-u\partial_x u
 \end{equation}
with $\eta\approx 0.075$.

Fig.\ \ref{fig_pattern}(b) presents another type of spatiotemporally
intermittent state that can be observed in the phase L2. Its
distinguishing feature is that it is composed of a set of virtually
immobile turbulent nuclei ``mediated'' by creation and absorption of
nonlinear waves, propagating trough the laminar background with unit
velocity. This type of ``nuclear'' STI state is characteristic for the
sub-phase $L2_n$. 

The phase L1 also features internal structure (see fig. \ref{fig_hyst}).
For $u_c<u<u_f$ (sub-phase $L1_l$) any initial configuration decays
quickly into a uniform stable state. In other words, the uniform state is
nonlinearly stable. For $u_f<u<u_d$ (sub-phase $L1_f$) the uniform state
is only conditionally stable and large deviations from it result in a
spatiotemporally chaotic asymptotic state. The transition point separating
the two sub-phases is estimated to be $u_f\approx 0.53$ in the
thermodynamic limit. 

Fig.\ \ref{fig_pattern}(d) represents a frozen pattern, characteristic of
the chaotic asymptotic state of the system in the sub-phase $L1_f$ and in
the phase T1 close to the boundary with L1 (we will denote this region
$T1_l$). It is probably more appropriate to call this type of dynamics
locked chaos: the chaotic state is almost stationary in terms of reduced
dynamics, with chaos localized in turbulent domains. As was briefly
mentioned in section \ref{sec_phase}, the dimensions and locations of the
turbulent domains depend on the details of the initial state of the
system. The laminar state in this sub-phase is uniform. 

And finally, fig. \ref{fig_pattern}(c) gives the reduced dynamics
representation of critical behavior displayed by the CML in the chaotic
phase T2 close to the boundary with L1. Here we encounter yet another
example of a spatiotemporally intermittent state observed in our model.
The laminar state is again a 2-cycle (more specifically the type-II
2-cycle) and it is also absorbing, i.e a new turbulent domain (we will
call these defects because of their small size) can never originate inside
a laminar domain, it can only be spawned by other defect(s). However a
defect can be consumed by a laminar domain, or alternatively it can be
destroyed in a collision with another defect. 

The most prominent feature of this picture is ``spontaneous'' creation and
annihilation of turbulent pulses (defects) moving in different directions
with different (but constant) velocities. Therefore we may alternatively
regard these defects as traveling waves. Naively one would expect that
the condition $\xi\gg 1$, where $\xi$ is the correlation length, is
necessary and might also be sufficient for the formation of a number of
traveling waves. Numerical results for our model support this assumption
(in disagreement with the stronger restriction \cite{ciliberto}, according
to which the correlation length $\xi$ should be comparable to the size of
the system $L$). 

Nevertheless, since in the strongly chaotic regime the correlation length
is of order one lattice spacing, the condition $\xi\gg 1$ is usually only
satisfied close to the hypersurfaces in the parameter space on which the
correlation length diverges, i.e. where a continuous phase transition
occurs. This is clearly the case of fig. \ref{fig_pattern}(c): we have a
continuous phase transition at $u=u_c\approx 0.4615$. 

Deeper in the chaotic phases T1 and T2, away from the phase boundaries,
strongly chaotic behavior could be observed. Here almost all sites on the
lattice exhibit turbulent behavior, and only occasionally a laminar domain
of a very small size is created and then quickly consumed by the
neighboring turbulent sites. We would call this type of dynamics strong
chaos in contrast to the mild chaos, where all chaotic dynamics is
localized to turbulent domains, occupying only a part of the lattice,
while the rest of it is in the laminar state. 

\section{Domain Lengths}
\label{sec_domain}

A quantitative description of the reduced dynamics is provided by the
probability distribution functions $P_t(l)$ and $P_l(l)$ giving the
probability for a turbulent (laminar) domain to have length $l$. In order
to calculate these functions numerically we used a single random initial
condition and let the system evolve, counting how many times a laminar
(turbulent) domain of a given size formed. The resulting distributions did
not depend on a particular choice of the initial condition for any given
set of control parameters corresponding to ergodic dynamics, i.e
everywhere except the sub-phase $L1_f$. 

Our calculations show that in the chaotic state (away from phase
boundaries) both $P_t(l)$ and $P_l(l)$ decay exponentially. Even more
important, typical lengthscales appear to be almost independent of the
value of the conserved quantity $u$, being $l_l\approx 1$ for laminar and
$l_t\sim 10^3$ for turbulent regions. This is consistent with the point of
view that most of the system is in a turbulent state and the probability
of encountering a laminar region at any particular location is very small
(but finite and independent of the system size) and decreases rapidly with
increasing domain length, while there is no spatial structures defining
alternative lengthscales. 

The behavior of length distribution functions might be expected to change
substantially as the system gets close to the phase transition points. For
example, if the transition is continuous the correlation length may grow
and the critical effects might introduce alternative lengthscales that
would modify the form of the distribution functions. 

We first examine the ``percolating'' STI state characteristic of the
sub-phase $L2_p$. Fig.\ \ref{fig_domain}(a) shows that both $P_t(l)$ and
$P_l(l)$ decay exponentially for large $l$. This is compatible with the
assumption that the phase transition at $u=u_p$ might in fact belong to
the universality class of directed percolation \cite{grassberger}. It is
interesting to note though, that $P_l$ has two branches, one corresponding
to even, the other to odd size of a domain. About the only useful
information that one can extract from this data is typical length scales
of laminar and turbulent domains, which, for $u=0.06$, appear to be tens
of lattice spacings for both. 

In the previous section we saw that the STI states observed inside $L2_p$
and $L2_n$ are considerably different. As a result fig.\
\ref{fig_domain}(b), which corresponds to the ``nuclear'' state, differs
from fig.\ \ref{fig_domain}(a) substantially: though $P_t(l)$ still decays
exponentially, $P_l(l)$ does not, but has another peak at $l_s\approx200$.
This peak is not a finite size effect (which can be shown using a larger
system) and indicates the presence of an internal spatial structure with
characteristic length $l_s$ (typical separation between the ``nuclei'') in
the STI state. The typical width of the ``nuclei'' is, in turn, determined
by $P_t(l)$. 

Numerical data for a larger system ($L=4096$) suggest (in contrast with
the results obtained by Chate and Manneville \cite{chat_man}) a crossover
type of behavior for the distribution of laminar domain lengths away from 
the onset of STI: 
 \begin{equation} 
 P_l(l)\sim\cases{ 
 l^\alpha,\quad \alpha<0 &\text{for $1\ll l\ll l_s$},\cr 
 exp(-l/l_s) &\text{for $l\gg l_s$}.} 
 \end{equation}
 As a result, if the characteristic length $l_s\rightarrow \infty$ as
$u\rightarrow u_n^{+}$ we should expect a pure power law decay of
$P_l(l)$ at $u=u_n$. 

Another interesting phenomenon, can be pointed out in fig.\
\ref{fig_domain}(c). Both $P_l(l)$ and $P_t(l)$ have two branches, one
corresponding to even, the other to odd length of a domain. The behavior
of these branches is quite peculiar, they cross at some crossover length
$l_{cr}\approx 36$: 
 \begin{eqnarray}
 P_l^{even}(l) > P_l^{odd}(l),\ l < l_{cr};\\
 P_l^{even}(l) < P_l^{odd}(l),\ l > l_{cr}.
 \end{eqnarray}
The difference between the branches is larger for $L$-even and smaller 
for $L$-odd. Similar relations hold for $P_t(l)$, though the difference 
between the branches is less pronounced than in the case of $P_l$.

Close to the boundary T2-L1 a typical state (fig.\ \ref{fig_pattern}(c))
is composed of a collection of laminar domains separated by small
turbulent defects. As we are going to see later, most of the turbulent
domains tend to have a fixed length $l_t=6$, independent of the distance
(in parameter space) to the transition point. 

All defects move with a constant velocity, but while the majority of
defects is moving with the maximal speed $v=\pm 1$ the rest have a smaller
speed $|v|\le 1$. Therefore we have a considerable probability of
encountering a laminar domain bounded by a pair of defects which move with
equal and opposite speeds. If at some particular time $n$ the length of
such a laminar domain was even (odd), it will remain even (odd) for as
long as these defects exist since the length will increase (decrease) by 2
at each time step. 

It is reasonable to assume that the details of the ``defect interaction''
favor the creation of domains with length of a given parity, say even over
odd. As a result, one will see that the probability of finding a laminar
domain of small and even size $l$ is higher than the probability of
finding a laminar domain with comparable odd size $l-1$ (or $l+1$), thus
splitting the function $P_l(l)$ into two branches. 

Since chaotic fluctuations tend to destroy the deterministic predictions
like the one we just discussed, we cannot make any rigorous conclusions
about the dynamics of larger laminar domains. The numerical data suggests
that the characteristic lengthscales determining the decay rates of the
two branches are different, so they cross at some crossover length
$l_{cr}$ determined by the average distances between the defects moving
with different speeds. 

In the case of a frozen pattern one should not expect the probability
density to decay exponentially, rather it should display several peaks,
broadened due to the chaotic fluctuations of the domain boundaries, at the
lengths occurring in a particular pattern and then go sharply to zero.
This applies to both $P_t$ and $P_l$. Fig. \ref{fig_domain}(d) satisfies
this prediction quite well. The underlying reason of this kind of behavior
is the non-ergodicity of the system in the frozen pattern forming regime. 

The numerical results in fig. \ref{fig_domain} imply that in the model
studied in this paper simply the ergodicity of the system dynamics is a
sufficient condition for the probability distribution function $P_t(l)$ to
decay exponentially at lengths greater than the correlation length $\xi$
in any disordered state.  On the contrary, the distribution of laminar
domains is more specific and informative. $P_l(l)$ still decays
exponentially in the strongly chaotic states. In the spatiotemporally
intermittent state however the behavior of $P_l$ varies widely: it might
or might not decay exponentially. For instance, in case of the ``nuclear''
STI state we observe the crossover from exponential to power law type of
decay. Similarly to \cite{chat_man} we expect to see a pure power law
decay at the transition point $u=u_n$. 

\section{Phase Transitions}
\label{sec_transition}

\subsection{Order Parameters}
\label{sec_oparam}

As mentioned above, we expect to have 4 distinct phase transitions, in
which the system goes from either uniform state or 2-cycle to a chaotic
state. The transition to chaos in this example of an extended system with
a local conservation law does not follow the period doubling cascade or
other routes to chaos characteristic of low dimensional dynamical systems,
instead the system goes directly from a simple dynamical state (fixed
point, period-2) to a chaotic state. In other words it has a character
similar to phase transitions in Hamiltonian statistical systems where a
symmetry of a basic state is destroyed upon crossing of the critical
point. This feature is common to all the phase transitions in this model. 

A conventional dynamical systems approach to the treatment of phase
transitions in a deterministic chaotic system would be to calculate the
maximal Lyapunov exponent, $\lambda_{max}$. The bifurcation from the
ordered to disordered state then occurs at the values of the parameters
where the exponent changes sign and becomes positive. In the case of CML
with a conservation law, one of the exponents is always zero, therefore
$\lambda_{max}$ does not change sign, but increases (continuously or
discontinuously, depending on the type of transition) from zero as the
system crosses the boundary between ordered and disordered phase. The
maximal Lyapunov exponent can be considered as an example of a global
(intensive) order parameter. 

On the other hand, in order to get some additional insight into the 
spatial dynamics of the system, it might be advantageous to introduce a
local (extensive) order parameter. A good candidate seems to be the 
density $h$ of the Kolmogorov-Sinai entropy, $S_{KS}$:
 \begin{equation}
 \label{eq_entropy}
 h={1\over L}S_{KS}={1\over L}\sum_{\lambda_m>0}\lambda_m.
 \end{equation}
It is clearly zero in the ordered phase and positive in the disordered
phase.

The calculation of the Lyapunov spectra is numerically costly, so a
different approach is used for comparison, based on the reduced
description of the system's dynamics. In terms of reduced dynamics the
phase transition from the ordered to disordered state can be represented
as the appearance of a set of disjoint turbulent domains on the laminar
background. This leads us naturally to the measure of the set of turbulent
domains, $\rho_t$, as another local order parameter, describing how the
laminar state becomes turbulent. In the ordered state $\rho_t=0$, in the
disordered state $0<\rho_t<1$. 

And, finally, let us introduce yet another order parameter, $e_{ch}$. One
can decompose the chaotic dynamics into modes using the Karhunen-Loeve
decomposition \cite{nicolaenko}. The mode intensity $E_k$ is defined as
the eigenvalue of the integral equation
 \begin{equation}
 \sum_{i=1}^L K(j,i)\psi_k(i) = E_k\psi_k(j),
 \end{equation}
where the kernel
 \begin{equation}
 K(i,j)=<u_i^n\,u_j^n>_n
 \end{equation}
 is just the 2-point correlation function. Due to the translational
invariance of the system (in case of ergodic dynamics) $K(i,j)=C(i-j)$ and
therefore the eigenfunctions $\psi_k(i)$ are just Fourier modes.
Consequently, the eigenvalues $E_k$ are given by the values of the static
structure function
 \begin{equation}
 E_k=S(k)=<|u^n_k|^2>_n,
 \end{equation}
where $u^n_k$ is the Fourier transform of the map variable.

The total intensity of the dynamics is defined as $E=\sum_m E_m$. It
includes the contributions from the chaotic as well the non-chaotic modes. 

In the non-chaotic phases L1 and L2 the stable stationary state is given
by the general formula (\ref{eq_zig}). Therefore, the structure function 
might only be nonzero at two values of the wave vector $k$:
 \begin{equation}
 S(0)=u^2
 \end{equation}
and, for a zig-zag state,
 \begin{equation}
 S(\pi)=A^2.
 \end{equation}

In other words, if we want the order parameter to represent the strength
of chaos in the system, it should be defined through the intensity of the
chaotic modes only. So we arrive at the following expression
 \begin{equation} 
 e_{ch}={1\over L}\sum_{m=1}^{{L\over 2}-1}S(k_m),\ \ k_m={2\pi m\over L},
 \end{equation}
 which is identically equal to zero in any ordered phase and larger than
zero in any disordered phase. 

All order parameters introduced are expected to become asymptotically
independent of the system size $L$ in the thermodynamic limit
$L\rightarrow \infty$.

\subsection{Continuous transition}
\label{sec_cont}

Most of the attention in this study was devoted to the order-disorder
transition occurring at the boundary L1-T2. This transition is easy to
study and it is expected to be quite common in models described by a
CML with diffusive coupling and conserved map variable density. 

The transition point is defined by the equation $f'(u_c)=1/2$. One can see
from eq.\ (\ref{eq_lambda}) that at $u=u_c$ a period 2 bifurcation occurs
and the $k=\pi$ mode becomes growing, making the uniform configuration
unstable. 

Details of the transition (see next section) suggest that this transition
is a continuous (second order) phase transition, and the numerical data
support this conclusion. One of the clear indications of this fact is
presented in fig.\ \ref{fig_corl}, which shows that the correlation length
diverges as we approach the phase transition point
 \begin{equation}
 \xi\propto (u_c-u)^{-\nu}
 \end{equation}
 with critical exponent $\nu$ estimated to be of order $0.8$. The
correlation length is hard to measure however, and the precision of this
result is low, so that this value cannot be considered reliable. 

A better diagnostic of a diverging lengthscale is the average length of a
laminar domain $l_l = \sum_l l P_l(l)$ which is seen to scale (fig.
\ref{fig_length}(a)) as
 \begin{equation}
 l_l\propto (u_c-u)^{-\mu},\qquad \mu\approx 1.0\pm 0.02
 \end{equation}
which implies, in particular, that very close to the transition point the 
lattice configuration consists of a few large laminar domains, separated by
turbulent defects of finite size.

The average length of a turbulent domain $l_t = \sum_l l P_t(l)$ does not
scale (fig.\ \ref{fig_length}(b)), but shows an exponential dependence on
the distance from the critical point: 
 \begin{equation}
 l_t\approx l_t^{cr}\ exp({u_c-u\over u_t}),\ \ u_t\approx 0.006,
 \end{equation}
approaching the limiting value 
 \begin{equation}
 \label{eq_turl}
 \lim_{u\rightarrow u_c}l_t=l_t^{cr}=6.
 \end{equation}
 This means that the onset of disorder at this particular phase transition
is dominated by the creation of defects of the fixed width $l_t^{cr}$. In
particular, one can check that the width of all the defects in fig.\
\ref{fig_pattern}(c) is about 6 lattice spacings.

All of the order parameters defined above take a zero value in the
non-chaotic phase and increase continuously from zero in the chaotic phase
as we move away from the transition point. What is more interesting, they
scale algebraically with the deviation of $u$ from the transition point
(fig.\ \ref{fig_oparam}), all with different critical exponents: 

 \begin{eqnarray}
 \lambda_{max}\propto (u_c-u)^{\beta_\lambda},
    \qquad \beta_\lambda\approx 0.8\pm 0.03; \label{eq_lam}\\
 h\propto (u_c-u)^{\beta_h},
    \qquad \beta_h\approx 2.0\pm 0.05; \label{eq_ent}\\
 \rho_t\propto (u_c-u)^{\beta_\rho},
    \qquad \beta_\rho\approx 1.0\pm 0.01; \label{eq_rho}\\
 e_{ch}\propto (u_c-u)^{\beta_e},\qquad \beta_e\approx 0.5\pm 0.01.
 \end{eqnarray}

One can use these values to compare the transition with similar phase
transitions in other conserving system, both continuous and discrete, and
perhaps determine whether this transition belongs to some universality
class. 

\subsection{Critical exponents}
\label{sec_crit}

A closer look at the details of the phase transition at $u=u_c$ reveals
that equation\ (\ref{eq_rho}) is to be expected and that the transition
should necessarily be continuous, as a result of the conservation law and 
a particular feature of the local map $f(x)$.

As was mentioned in the previous section, the change in the growth rate
of the $k=\pi$ mode is responsible for the transition. One can determine
from eq.\ (\ref{eq_lambda}) that for $u$ close to $u_c$, equating the
Lyapunov exponent with the growth rate, the linear stability analysis 
of the uniform state gives
 \begin{equation}
 \label{eq_maxexp}
 \lambda_{\pi}\approx -8b(u-u_c)
 \end{equation}
 which changes sign as the system moves across the transition point, from
the ordered state L1 to disordered state T2. At the transition point the
growth rate obviously vanishes, making the zig-zag state neither stable
nor unstable in the linear sense. In fact as mentioned in section\
\ref{sec_phase}, the zig-zag state given by eq.\ (\ref{eq_zig}) (type-II
2-cycle) with arbitrary amplitude $A$ can exist at $u=u_c$ (and only at
$u=u_c$) and is stationary, meaning that the amplitude $A$ neither grows
nor decays. This is in contrast with the result for phase L2 (type-I
2-cycle), where the amplitude of the stable state is defined by eq.\
(\ref{eq_ampl}). 

This fact results in some interesting consequences for the system dynamics
in the disordered phase close to the transition point. Most of the lattice
develops a zig-zag pattern (fig.\ \ref{fig_lattice}) similar to the one we
just discussed
 \begin{equation}
 u^n_i=u_c+(-1)^{i+n} A_i,
 \end{equation}
 where now the amplitude $A_i$ is not a constant, but a slowly varying
function of the lattice site. The whole lattice cannot be in such a state
for $u<u_c$ because of the conservation law. In order to compensate for
the difference, several similar turbulent defects separating laminar
domains form, with a fixed width $l=l_t^{cr}$ (see previous section) and
the local density of the map variable
 \begin{equation}
 {1\over l}\sum_{i=i_0}^{i_0+l-1} u_i^n = u_c-\delta u
 \end{equation}
 which is lower than the critical value $u_c$ by $\delta u$. 

Here $\delta u$, should not strongly depend on $u$, because $u$ is not a
local parameter prescribing the dynamics. On the contrary, the local
density in the turbulent defect is only determined by the structure of the
interface separating two laminar domains that have their local densities
fixed at $u_c$, independent of $u$. The structure of the interface, in
turn, depends primarily on the width of the turbulent domain, which is
seen to depend very weakly on $u$. Numerical results support this
conclusion. 

This results in the value of the conserved quantity being ``adjusted'' to
comply with the conservation law to give on average
 \begin{equation}
 \rho_l u_c+\rho_t(u_c-\delta u)=u.
 \end{equation}
 Now we can easily extract the dependence of $\rho_t$ on $u$. Since 
$\rho_t+\rho_l=1$
 \begin{equation}
 \rho_t={(u_c-u)\over \delta u}.
 \end{equation}

This derivation confirms the value of the critical exponent $\beta=1$. 
Thus the conservation law is ultimately responsible for the way this
particular phase transition occurs and for its type. The ordered state
turns into a disordered one by developing a set of very similar turbulent
domains (defects), which have a fixed length (the deviation in eq. 
(\ref{eq_turl}) from the $l_t=6$ is due primarily to ``defect
interaction'' effects), but whose number increases as the system moves
further and further away from the transition point, so as to compensate
for the change in the density of the map variable in the laminar regions
with respect to the average value given by the conservation law. We may
therefore suggest that defects play a more important role in
order-disorder transitions in conserving systems than in non-conserving
systems \cite{chat_man_2}. 

The value of the critical exponent for the maximal Lyapunov exponent
$\beta_\lambda\approx 0.8$ numerically obtained in eq.\ (\ref{eq_lam}) is
different from the one predicted by eq.\ (\ref{eq_maxexp}). However the
latter value is calculated for a reference trajectory corresponding to the
uniform configuration and since for $u<u_c$ the system is in the chaotic
state, the validity of eq.\ (\ref{eq_maxexp}) is far from being obvious,
no matter how small the distance to the transition point is.  We should
also consider finite size effects. It is natural to expect that numerical
and theoretical values agree if there is just one positive Lyapunov
exponent and therefore no mode mixing. From eq.\ (\ref{eq_lambda}) it
follows that the next mode to become growing is the mode with
$k=\pi-{2\over L}\pi$ and this happens when
 \begin{equation}
 u=u_c-\Delta u,\ \ \Delta u\approx {\pi^2\over 16bL^2}.
 \end{equation}
So we might expect the crossover behavior for the exponent $\beta_\lambda$:
 \begin{eqnarray}
 \label{eq_crover}
 \beta_\lambda\approx 0.8,\ \ u_c-u>\Delta u,\\
 \beta_\lambda=1.0,\ \ u_c-u<\Delta u.
 \end{eqnarray}
 In practice the value of $\Delta u$ was usually so small ($\Delta
u\approx 3\cdot 10^{-5}$ for $L=128$), that eq.\ (\ref{eq_crover}) was
satisfied for all deviations of $u$ from the critical value used in our
numerical calculations (at most $2\cdot 10^{-4}$). 

And finally we would like to mention that the value of the critical
exponent $\beta_h$ can be evaluated from the limiting form of the Lyapunov
spectrum. Close to the transition point the positive Lyapunov exponents
are well approximated by the following expression: 
 \begin{equation} 
 \lambda_m=\lambda_{max}-c_1 m^\gamma, \qquad \gamma=0.7\pm0.1 
 \end{equation}
 Together with the equations (\ref{eq_ent}) and (\ref{eq_lam}) this
implies that the critical exponent corresponding to the Kolmogorov-Sinai
entropy density is given by
 \begin{equation}
 \beta_h=\beta_\lambda\,(1+{1\over\gamma}),
 \end{equation}
that yields the value $\beta_h=1.94\pm0.20$ consistent with eq.\ 
(\ref{eq_ent}).

We might expect these exponents to define a universality class for the
onset of spatiotemporal chaos. The arguments leading to the predictions
for the values suggest that the class may depend both on the existence of
a conservation law and on special symmetry properties of the map function.
The restrictions on $f(x)$ can be obtained in the following way. 

One starts with the relation between the amplitude and the local density
of the type-II 2-cycle for an arbitrary $f(x)$: 
 \begin{equation}
 \sum_{n-odd}{\frac{1}{n!}}A^{n-1}f^{(n)}(u)={\frac{1}{2}}\ .
 \end{equation}

Close to the transition point $u=u_c$ it can be rewritten as 
 \begin{equation}
 0=-r(u-u_c)A+d_3A^3+d_5A^5+\cdots
 \end{equation}
where $r=f^{\prime\prime}(u_c)$ and coefficients $d_n$ are defined as 
 \begin{equation}
 d_n={\frac{1}{n!}}f^{(n)}(u_c),\qquad n\ge 3.
 \end{equation}

Depending on the values of parameters $r$ and $d_{3}$, we can have either
a subcritical or supercritical bifurcation at the transition point. The
special case studied in this paper corresponds to
 \begin{equation}
 d_{3}=d_{5}=\cdots =0  \label{eq_cond}
 \end{equation}
and is intermediate between the two types of bifurcation. 

These special properties follow immediately for continua of $a$ and $b$
from the parabolic nature of the map we have used. More generally we can
consider maps of the form
 \begin{equation}
 f(x)=ax+bs(x),  \label{eq_general}
 \end{equation}
 where $s(x)$ is an arbitrary function symmetric about its maximum. By
rescaling and shifting the origin of $x$ and choice of normalization of
$s$ we can set $s(0)=0$, the maximum of $s$ to occur at $s=1/2$ and then
$s(1/2)=1$, leaving the two parameters $a$ and $b$ as well as the
conserved quantity $u$ to define the system. For this general family of
maps the degenerate bifurcation to the period-$2$ state occurs only for
$a=1/2$ and for $u$ at the maximum of $s$ i.e. $u=1/2$. Thus the
universality class is codimension-2 --- two parameters must be tuned to
arrive at this type of transition. 

For other values of $a$ and $b$ in (\ref{eq_general}) the bifurcation to
the period-2 state will be either supercritical or subcritical. For the
supercritical case a stable laminar 2-cycle state develops, which may be
the first step in a subharmonic cascade. For the subcritical case
attractors develop far away in phase space, and a full non-linear analysis
is needed to determine the type of behavior. 

\subsection{Hysteretic transitions}
\label{sec_hysteresis}

Now we turn our attention to the phase transitions that we expect to occur
at the boundaries L2-T1, L2-T2 and L1-T1. As we are going to see later,
all three are very similar, so we will concentrate on the transition at 
L2-T1 below. 

There is a considerable difference between the transition at L1-T2 and the
transition at L2-T1: in the former case the asymptotic state in both L1
and T2 is unique, while in the latter case the asymptotic state in the
ordered phase L2 can be either ordered, or spatiotemporally chaotic. As a
result we should specify between which states the transition occurs.

It will be convenient to introduce an additional parameter $v(u)$
characterizing the volume of the of the basin of attraction. For example,
as we know from section \ref{sec_domain} in the sub-phase $L2_l$
($u_p<u<u_n$) the attractor is unique and therefore the basin of
attraction is the whole configuration space with volume
$v_l(u)=1^{L-1}=1$. 

In the sub-phase $L2_p$ the non-chaotic attractor coexists with the
chaotic one, so we have $0<v_l(u)<1$ for $u_a<u<u_p$. Since there are no
other attractors in this sub-phase, the volume of the basin of attraction
of the chaotic attractor is given by $v_t(u)=1-v_l(u)$. Numerical data
suggest that most of the initial conditions in $L2_p$ result in
spatiotemporally intermittent chaotic asymptotic state and therefore
typically $v_t(u)\gg v_l(u)$. Moreover $v_l(u)\rightarrow 0$ as
$u\rightarrow u_a^{+}$ while $v_t(u)\rightarrow 0$ as $u\rightarrow
u_p^{-}$. 

At the point $u=u_a$ the 2-cycle state loses its stability through a
subcritical bifurcation. Outside the phase L2 the 2-cycle state cannot
exist and as the system crosses the phase boundary all order parameters
jump from zero in the phase L2 to some nonzero values in the phase T1 (see
fig. \ref{fig_hyst}). As a result one observes a discontinuous transition
in the non-chaotic state: the laminar state abruptly turns into the
chaotic one. 

However, there is apparently no phase transition in the chaotic state at
this point:  the dynamics of the system on the T1 side of the boundary is
very similar to the dynamics of the system evolving on the chaotic
attractor on the L2 side. All order parameters change continuously as the
system crosses the boundary from L2 to T1. Since $v_l(u)\rightarrow 0$ as
$u\rightarrow u_a^{+}$, changing the direction does not modify this
conclusion: the chaotic attractor of the phase T1 smoothly transforms into
the chaotic attractor of the phase L2. 

At $u=u_p$ the situation is reversed. Obviously there could be no phase
transition in the non-chaotic state. On the other hand, the chaotic
attractor does not exist in $L2_l$, therefore there should be some kind of
phase transition in the chaotic state: the order parameters are zero for
$u_p<u<u_n$, but take on nonzero values in the chaotic state for
$u_a<u<u_p$. Thus $u=u_p$ corresponds to the onset of STI.

The coexisting attractors form a hysteresis loop in the sub-phase $L2_p$
(fig. \ref{fig_hyst}). If we start at $u>u_p$ and gradually decrease
parameter $u$ the system will remain in the non-chaotic state while
$u>u_a$ and then jump to the chaotic one at $u=u_a$. Conversely starting
at $u<u_a$ and gradually increasing parameter $u$ makes the system remain
in the chaotic state while $u<u_p$. At $u=u_p$ the chaotic state becomes
non-chaotic thus closing the loop. 

There is a numerical complication here: it is not possible to establish
the exact value of the critical parameter $u_p$ for a system of finite
size. For $u\rightarrow u_p^{+}$ the lifetime of the chaotic transients
becomes very long and one cannot reliably determine the type of the
asymptotic state. On the other hand, $v_t(u)\rightarrow 0$ as
$u\rightarrow u_p^{-}$, which means that it becomes increasingly hard to
find initial conditions resulting in a persistent chaotic state as $u$
gets close to $u_p$, especially for small systems. Although $v_t(u)$ grows
rapidly with the system size, the smaller $|u-u_p|$ is the larger the
system should be in order to obtain the persistent chaotic state.
Therefore one cannot reliably determine the values of the order parameters
in the chaotic state close to $u=u_p$. 

As a result, it is even hard to determine reliably whether the transition
at $u=u_p$ in the chaotic state is actually continuous, although the
correlation length seems to diverge approaching the transition point. The
order parameters do not provide a clear picture either. Both $h$ and
$\rho_t$ seem to increase continuously from zero, but $\lambda_{max}$ and
$e_{ch}$ jump discontinuously as the system moves across the transition
point to the chaotic sub-phase $L2_p$.  All order parameters gradually
increase as the system moves away from the transition point toward the
chaotic phase T1. The number of positive Lyapunov exponents also grows
showing the increase in the number of chaotic modes. These results are
consistent with \cite{chate}. But so far we do not have reliable data that
will allow us to determine whether this transition in fact belongs to the
universality class of directed percolation or not. 

There is a similarity between the transition at $u=u_p$ and the continuous
phase transition at $u=u_c$: the laminar state becomes chaotic through the
appearance of a set of turbulent domains that gradually spread over the
whole system. Nevertheless there is an important difference: the STI state
observed inside T2 can never become completely laminar because of the
conservation law, while the STI state inside $L2_p$ may decay into a
completely laminar state with time. 

The above discussion applies completely to the chaotic sub-phase $L2_n$
which corresponds to $u_n<u<u_b$. One just has to replace T1 with T2,
$u_p$ with $u_n$ and $u_a$ with $u_b$. In particular, $v_l(u)\rightarrow
0$ as $u\rightarrow u_b^{-}$. One particular feature of the chaotic
attractor in this sub-phase is worth mentioning: for $0.1\lesssim u
\lesssim 0.15$ the measure of the turbulent set, $\rho_t$ grows linearly
with $u$. 

The phase transition L1-T1 at $u=u_d$ is also very similar to the
transition L2-T1. One just has to replace $L2_p$ with $L1_f$,
$u_p$ with $u_f$, and $u_a$ with $u_d$ in the above discussion.

Here, unlike the phase L2, the laminar state is the uniform state, not the
2-cycle. The uniform state is unstable for $u>u_d$, so in order for the
laminar regions to be stable the local density inside them should be in
the range $u_c<u<u_d$ characteristic of the ordered phase L1. This is in
fact the case: the value of the local density in the laminar regions is
$u_l\approx 0.53$. 

Inside $L2_f$ the number of positive Lyapunov exponents grows linearly
with $\rho_t$, while the latter grows linearly with $u$. This is the kind 
of behavior one expects in a system where all chaos is localized in 
turbulent domains. 

All three cases of hysteretic transitions observed in our model are very
similar and represent many of the characteristic features of the specific
route to chaos. In particular, STI appears whenever the uniform absorbing
state experiences a subcritical bifurcation at the transition point (e.g.
$u=u_a$); the hysteresis loop forms as a result of the bistability; onset
of STI (e.g at $u=u_p$) can be either continuous or discontinuous,
depending on the types of defects supported by the laminar (absorbing)
state; the lifetime of the ``laminar'' transients diverges at the onset;
the confinement effects enhance the stability of the absorbing state in
relatively small systems. 

\section{Lyapunov Dimension}
\label{sec_lyapunov}

In this section we will briefly comment on the applicability of the
Lyapunov dimension as a parameter characterizing the dynamics in the
system under consideration. This has been used by a number of authors
(e.g. \cite{manneville_2,ruelle}) to assess the strength of chaos in
nonlinear dynamical systems. 

The Lyapunov dimension is defined as
 \begin{equation}
 D_L=n+{\nu_n \over \nu_n + \nu_{n+1}},
 \end{equation}
 where $\nu_n=\sum_{i=1}^n \lambda_i$ and $n$ is such that $\nu_n>0,
\nu_{n+1}<0$. It was suggested in ref. \cite{cross} that $D_L$ is not
defined for the system considered, because for the values of the control
parameters used all $\nu_n$'s were positive. 

This conclusion appears incorrect. First of all, this only happens for
some restricted set of control parameters. For other values of parameters
the Lyapunov dimension is perfectly well defined. Second, there is no
problem with all $\nu_n$'s being positive: it only means that $D_L = L-1$,
i.e. the dimension of the attractor coincides with the dimension of the
configuration space ($L$ variables with one constraint) meaning that the
attractor fills the configuration space. As one can see from the fig.
\ref{fig_dimension} this only happens away from the boundaries inside the
chaotic phases T1 and T2, where the system is in the strongly chaotic
regime with $\rho_t\approx 1$. 

The numerical results show that the Lyapunov dimension thus defined is a
good (albeit costly to calculate) measure of the strength of chaos in the
system under consideration. It is a continuous function of control
parameters and can in principle be used as an alternative order parameter,
though it proved to be hard to calculate with the necessary precision
close to the phase transition points. 

\section{Singularities in the Lyapunov Spectrum}
\label{sec_singular}

\subsection{Introduction}
\label{sec_sintro}

It has been suggested \cite{bohr} that all coupled maps with a
conservation law should have a singularity in the spectrum of Lyapunov
exponents at the value $\lambda=0$. The origin of the divergence of the
number of Lyapunov modes with negative exponents close to zero is
generally explained by the following arguments. 

A spatially uniform autonomous system with a locally conserved density is
considered. The sufficiently coarse grained asymptotic state of such a
system is supposed to be uniform, i.e the density in the asymptotic state
should not depend on the spatial coordinates. Now a long wavelength
density perturbation is imposed on such an asymptotic state. Because of
the conservation law the only mechanism by which such a perturbation can
decay is diffusion from the regions with high average density to the
regions with low average density. This process can be described by an
effective diffusion equation, which can be considered as a long wavelength
approximation of the original evolution equation. As a result the decay 
rate of a long wavelength perturbations should be given by a quadratic 
function of the wave vector $k$.

In our model this functional dependency can be motivated by analogy with
the analytic expression obtained for a stable non-chaotic phase L1. Since
there is an exact 1-to-1 correspondence between Fourier mode number and
the decay rate in L1, we can formally rewrite eq.\ (\ref{eq_lambda}),
re-expressing the number of the exponent $m$ through the wave vector $k$: 
 \begin{equation}
 \lambda (k)=\ln|1-4(a+b-2 bu)\sin^2({k\over 2})|.
 \end{equation} 
 Here for $k\rightarrow 0$ we have $\lambda (k)\propto k^2$ and the density
of Lyapunov exponents diverges
 \begin{equation}
 \label{eq_diverge}
 n(\lambda)\propto |\lambda|^{-1/2}\rightarrow\infty,\qquad 
 \lambda\rightarrow 0^{-}. 
 \end{equation}

One can attempt to apply a similar numerical analysis in the chaotic
phase. We start with writing the evolution equation (\ref{eq_cml}) in
Fourier space: 
 \begin{equation} 
 \label{eq_cml_f}
 u_k^{n+1}=u_k^n-4(a+b-2bu)\sin^2({k\over 2})u_k^n+
 4b\sin^2({k\over 2})\sum_{k'\ne 0,k}u_{k'}^n u_{k-k'}^n. 
 \end{equation}
 Consequently the Jacobian $J^n_{lm}=\partial u^{n+1}_l/ \partial u^n_m$
of the evolution transformation takes on the following form in the Fourier
space:
 \begin{equation}
 \label{eq_jacob}
 J^n_{kk'}=(1-4(a+b-2bu)\sin^2({k\over 2}))\,\delta_{kk'}+
           8bu^n_{k-k'}\sin^2({k\over 2})(1-\delta_{kk'}).
 \end{equation}
 Fourier modes are not eigenvectors of this matrix (unless all $u^n_k=0$
as in the uniform state), but since the off-diagonal elements are of order
$O(k^2)$ Fourier modes can serve as a good first order approximation to
the exact eigenvectors for $k$ sufficiently small. This argument suggests
that wavevector $k$ should be a good label for the slow modes (and small
$\lambda$) of the system in chaotic as well as spatially ordered states. 

\subsection{Structure of the Lyapunov Vectors} 
\label{sec_}

Numerically the Lyapunov spectrum can be calculated using the QR
decomposition of the product of the Jacobians $J^n_{lm}$
 \begin{equation}
 J^nQ^{n-1}=Q^nR^n,
 \end{equation}
where $Q$ is orthogonal and $R$ -- upper-diagonal to yield:
 \begin{equation}
 J^n\cdots J^0=Q^nR^n\cdots R^0=Q^n\tilde{R}^n.
 \end{equation}
 The columns of $Q^n$ give Lyapunov vectors and the diagonal elements of
$\tilde{R}^n$ -- the corresponding Lyapunov exponents on the n-th step: 
 \begin{equation}
 \lambda_{m}={1\over n}\ln(\tilde{R}^n_{mm})=<\ln(R^n_{mm})>_n.
 \end{equation}

Figs.\ \ref{fig_spec_70}-\ref{fig_spec_455} show several typical time
averaged power spectra of the instantaneous Lyapunov vectors along with
the corresponding Lyapunov spectra. The power spectra are represented in
the form of the density plots showing the relative contribution $P_m(k)$
from the Fourier mode with number $kL/2\pi$ to the $m$-th Lyapunov vector,
while the Lyapunov spectra show the correspondence between Lyapunov
vectors and exponents. The power spectra are normalized
 \begin{equation}
 \sum_k P_m(k)=1,
 \end{equation}
 so that $P_m(k)$ can alternatively be interpreted as the probability
distribution functions. 

We would like to mention that due to the ergodicity of the dynamics the
power spectra (as well as Lyapunov spectra) should be reproducible, i.e. 
different initial configuration with the given set of control parameters
should produce the same spectrum and this is found to be the case. It is
interesting to note though, that we obtained a unique form for the spectra
even for the values of the control parameters producing frozen patterns,
where we expect ergodicity to break down. 

We start with the spectra calculated inside the chaotic phases T1 and T2.
One can easily notice that a singularity appears in the spectrum of
Lyapunov exponents (fig.\ \ref{fig_spec_70}(b)) when the dominant
contribution to the Lyapunov vectors, corresponding to a slow evolution
(small $\lambda$) comes from the long wavelength Fourier modes and there
is at least an approximate 1-to-1 correspondence between the exponent
number and the dominant Fourier mode number for the relevant vector (fig.\
\ref{fig_spec_70}(a)). It is therefore natural to expect the small
negative Lyapunov exponents to be determined by the decay rate of the
corresponding Fourier mode. 

In the long-wavelength limit $k\rightarrow 0$, the latter could be
determined from a hydrodynamic analysis \cite{halperin} of the problem. It
was argued \cite{cross} that at long wavelengths the effect of all short
wavelength corrections to the equation of motion (\ref{eq_cml_f}) can be
combined into a stochastic noise term and a renormalized diffusion
constant $D$, producing an effective (discrete) Langevin equation: 
 \begin{equation}
 \label{eq_flang}
 u_k^{n+1}-u_k^n=-D k^2 u_k^n + k^2 \eta_k^n
 \end{equation}
 with $\delta$-correlated noise term
 \begin{equation} 
 <\eta_i^n\eta_{i'}^{n'}>=B\delta_{ii'}\delta_{nn'}.
 \end{equation}
 According to the Central Limit Theorem such noise averages to small
values on large lengthscales and therefore, for small $k$, the decay rate
of the Fourier mode $u_k$ is equal to $-Dk^2$. 

Because of the conservation law one of the exponents in our model is
always equal to zero. Let it be $\lambda_{m_0}$. It obviously corresponds
to the Lyapunov vector with $k=0$. Typically, for small $\lambda$, the
relation between the number $m$ of the Lyapunov vector and the dominant
long-wavelength Fourier mode $k$ can be represented by the following
simple form: 
 \begin{equation} 
 \label{eq_dominant}
 k_m=\alpha {2\pi\over L}(m-m_0).
 \end{equation}
 Parameters $m_0$ and $\alpha\sim 0.5$ are determined from the Lyapunov
spectrum and power spectrum, respectively ($m_0=0, \alpha=0.5$ inside L1). 

The Lyapunov exponent $\lambda_m$ is in fact calculated as a time averaged
decay rate of the corresponding instantaneous Lyapunov vector. Since the
time averaged power spectrum $P_m(k)$ gives the averaged relative
contribution of the Fourier mode $k$ to the $m$-th Lyapunov vector, we
might estimate the value of the $m$-th exponent as
 \begin{equation}
 \label{eq_estimate}
 \lambda_m=-D\sum_k P_m(k)\,k^2.
 \end{equation}

As one can deduce from fig. \ref{fig_spec_70}(a), for $u=0.7$, $P_m(k)$ is
sharply peaked at $k_m$, so approximating $P_m(k)=\delta_{k_mk}$ we
readily obtain that the form of the Lyapunov spectrum for small negative
values of the exponent should be given by
 \begin{equation}
 \label{eq_spectrum}
 \lambda_m\approx -Dk_m^2,\qquad m=m_0,m_0+1,\dots
 \end{equation}
 One can easily check that the numerically calculated exponents are in 
fact given quite precisely by this expression with $D\approx 0.32$ 
for $0<k\lesssim\pi/4$. So we recover eq.\ (\ref{eq_diverge}), but now for 
the chaotic state.

Fig. \ref{fig_spec_30} provides us with another typical example of the
spectra corresponding to the strongly chaotic dynamics, now calculated for
$u=0.3$. One can notice that the singularity in Lyapunov spectrum in fig.
\ref{fig_spec_30}(b) is still present, although the quadratic fit provided
by (\ref{eq_spectrum}) is quite poor compared to the one for $u=0.7$. 

The power spectrum (fig. \ref{fig_spec_30}(a)) shows that, similarly to
the previous case, the dominant contribution to the Lyapunov vectors
corresponding to slow evolution comes from the long-wavelength modes, but
now the distribution $P_m(k)$ is much broader. This means that we can no
longer approximate $P_m(k)$ by $\delta_{k_mk}$ and there could be
considerable corrections to eq. (\ref{eq_spectrum}), which does not
however change the general conclusion about the presence of the
singularity in the spectrum of Lyapunov exponents at $\lambda\approx 0$. 

These arguments work well in the area of strong chaos, where the large
scale dynamics is determined by the long wavelength modes, i.e. modes with
$k\ll 1/\xi$, where $\xi$ is the correlation length in the system.  When
we approach a continuous phase transition, though, the correlation length
grows and becomes comparable to the system size $L$. When this happens,
the nonlinear terms in eq. (\ref{eq_cml_f}) become relevant on every scale
\cite{cross} and their effect can no longer be emulated by an effective
noise term. This results in a strong coupling between different long
wavelength modes (i.e. modes with $k\sim 1/L$) and therefore the
approximate 1-to-1 correspondence between the mode number and its growth
rate is completely destroyed. This means that there is no reason to expect
the divergence of the number of slowly evolving modes anymore. 

Fig.\ \ref{fig_spec_455}(a) suggests that there is a strong mode mixing in
the system and it is not possible to extract the dominant contribution to
the Lyapunov vectors. In fact, we really have a continuous phase
transition at $u=u_c$, very close to $u=0.455$, and it is responsible for
the disappearance of a singularity in the spectrum in fig.\
\ref{fig_spec_455}(b). One might still hope to find a trace of a
singularity at this particular value of $u$ by going to much larger system
size (higher resolution). 

\subsection{Positive side singularity}
\label{sec_posside}

Some models with a conservation law \cite{bohr} are found to have a
density of exponents diverging on the positive $\lambda$ side
 \begin{equation}
 n(\lambda)\rightarrow\infty,\qquad \lambda\rightarrow 0^{+}, 
 \end{equation}
 although this feature is not considered very common. This divergence is
present in our model too (see fig. \ref{fig_spec_70}(b) for example), but
is somewhat weaker than the divergence at negative $\lambda$: for some
values of the parameters it can be barely seen even for large lattices
(typically $L>10^2$ is necessary). For $\lambda\rightarrow 0^{+}$ the
spectrum can usually be fitted quite precisely by a quadratic function
 \begin{equation}
 \label{eq_pfit}
 \lambda_m\approx \tilde{D}(m-m_0)^2,\qquad m=m_0,m_0-1,\dots
 \end{equation}
 which means that the singularity of $n(\lambda)$ is inverse-square-root
on both positive and negative sides. 

Numerically calculated spectra show no sign of smoothing out of the
singularity (on either side) with increasing resolution (increasing size
of the system) up to $L=512$ (the results for $L=128$ and $L=512$ are
presented in fig. \ref{fig_cube}(a)) and also suggest that the fit
(\ref{eq_pfit}) could be good for positive $\lambda$ as large as
$0.3\lambda_{max}$. 

Figs. \ref{fig_spec_70}(a) and \ref{fig_spec_30}(a) suggest that it is
possible to extract the dominant Fourier modes corresponding to small
positive $\lambda$. Again their wavevectors scale roughly linearly with
$m_0-m$ and therefore equation (\ref{eq_dominant}) should be replaced by
 \begin{equation} 
 k_m=\tilde{\alpha} {2\pi\over L}(m_0-m)
 \end{equation}
 for $m<m_0$. Equation (\ref{eq_estimate}) though should be abandoned in
favor of a more precise one, preferably derived directly from the
evolution equation (\ref{eq_cml_f}). We intend to explore this question in
more detail later. 

\subsection{Effective diffusion constant}
\label{sec_diffconst}

We should also mention that the values of the effective diffusion constant
$D$ numerically obtained from eq. (\ref{eq_spectrum}) do not coincide with
the ones obtained through the dynamic structure function $S(k,t)$.
According to (\ref{eq_flang}) it is defined (see \cite{cross}) as
 \begin{equation}
 \label{eq_dsfun}
 S(k,t)=Bk^2e^{-Dk^2t}.
 \end{equation}
 It will be convenient to distinguish these using the notation $D_{ls}$
for the former and $D_{sf}$ for the latter. In the regions of parameter
space where the effective Langevin equation is applicable (and hence eq.
(\ref{eq_dsfun}) is valid) these approaches can give substantially
different results. This is yet another indication that equation
(\ref{eq_estimate}) can only provide a very crude estimate and should be
amended considerably to obtain adequate results. 

It is instructive to compare how $D_{ls}$ and $D_{sf}$ change if parameter
$b$ is varied while both $a$ and $u$ are fixed (see Fig. \ref{fig_diff}). 
For $u \gtrsim 1.3$ $D_{ls}>D_{sf}$ and they both grow with increasing
$b$. Numerical data available so far suggests that $D_{ls}/D_{sf}
\rightarrow 1$ as $b\rightarrow \infty$ (for strongly chaotic systems). 

For $u \lesssim 1.3$ $D_{sf}$ drops almost to zero indicating that there
is almost no diffusion in the system. Indeed we know that this is the
region of locked chaotic dynamics ($T1_l$). So this result is not
surprising: the formation of a locked structure prevents diffusion (on any
scale larger than some typical scale determined by that structure). We
cannot probe smaller scales using the dynamic structure function because
the effective Langevin equation (\ref{eq_flang}) is not valid for $k$
large enough, but supposedly diffusion survives there. It is interesting
to note however that locking has apparently no effect on the Lyapunov
spectrum: the change in $D_{ls}$ is very gradual across $T1$. 

\section{Role of the conservation law}
\label{sec_conserv}

Now that we have studied the dynamics of the CML with the conservation 
law quite thoroughly and compared the characteristic phenomena with those 
observed in CML's without any conservation laws we would like to discuss 
whether the distinguishing features are really explained by the presence 
of the conservation law.

It is often very hard to distinguish the effect of the conservation law on
the dynamics of a system from the effects introduced by other aspects of
the evolution equation. Here we would like to explore some consequences of
violating the conservation law, trying to retain the structure of the
original equation\ (\ref{eq_cml}). 

First of all, we are looking for internal homogeneous perturbations that
would violate the conservation law of eq.\ (\ref{eq_cml}), but would
preserve the structure of the equation. This is easily furnished by the
following modification of the original evolution equation: 
 \begin{equation}
 \label{eq_cml_n}
 u_i^{n+1}=u_i^n+\epsilon g(u_i^n)+(f(u_{i-1}^n)-2f(u_i^n)+f(u_{i+1}^n)),
 \end{equation}
 where the local map function $f(x)$ is the same as above and a
perturbation $g(x)$ is introduced. One can easily notice that in the
simplest case of the uniform state it reduces to the equation determining
the evolution of the average density
 \begin{equation} 
 \label{eq_1dmap} 
 u^{n+1}=u^n+\epsilon g(u^n). 
 \end{equation}
 This equation, though, does not provide us with any reliable information
concerning the dynamics of the average density in the case of a nonuniform
state. 

Second, in order to be able to compare two systems, one with and one
without the conservation law, we should ensure that the latter is violated
only ``mildly''. In other words, we would like the perturbed system to
have a phase diagram that could be compared to that of the original
system. In particular we would like to preserve the dimensionality of the
parameter space. Since $u$ is no longer conserved it is not a parameter of
the dynamics. Instead we introduce another parameter $u_0$ that will enter
the evolution equation through the perturbation function $g(x)$.

By ``mild'' conservation violation we also mean that the change of the
originally conserved average density $u$ during any single time-step is
sufficiently small: 
 \begin{equation}
 |u^{n+1}-u^n|\ll 1.
 \end{equation} 
We would also like the fluctuation of the average density $u$ to be
bounded, such that
 \begin{equation}
 u_0-\delta u<u^n<u_0+\delta u
 \end{equation}
for some finite $\delta u$ at any time step $n$. Then we would be 
able to compare the dynamics of the perturbed system with that of the 
original system with conserved quantity $u\approx u_0$.

Our primary interest in this section is the relation of the conservation
law to the existence of a singularity in the spectrum of Lyapunov
exponents. It is thus reasonable to compare the perturbed system with the
original one, described by the equation\ (\ref{eq_cml}), at the value of
the conserved quantity $u=0.8$, where the conserving system displays a
strongly chaotic dynamics and has a Lyapunov spectrum with a pronounced
singularity at $\lambda=0$. 

We will start with the following choice of the (nonlinear) perturbation
function: 
 \begin{equation} 
 \label{eq_nonlin}
 g(x)=(u_0-x)^3 
 \end{equation}
 and choose $\epsilon=0.2$. The dynamics of the perturbed system seems
qualitatively very similar to the dynamics of the unperturbed system, but
now the average density $u$ is not conserved and fluctuates about $u_0$
with a standard deviation of order few percent. 

Fig. \ref{fig_cube} of the Lyapunov exponent spectra focuses on the parts
corresponding to slow evolution (small $\lambda$). Comparing the spectrum
of the modified system with that of the original system for $u=u_0=0.8$,
one can easily notice that the singularity is clearly present in the
conserving case, while the spectrum of the perturbed system appears to be
similar to the spectrum of the original system, but slightly tilted and
shifted downwards. 

In order to better understand the origin of such a metamorphosis it is
advantageous to use another type of perturbation which is a lot easier to
interpret and study analytically:
 \begin{equation} 
 \label{eq_dissip}
 g(x)=u_0-x. 
 \end{equation} 
 Since $g(x)$ is linear, equation\ (\ref{eq_1dmap}) now describes the
evolution of the average density $u$ for arbitrary initial state. For
$\epsilon>0$ ($0.001\le\epsilon\le 0.1$ was used) the asymptotic state is a
configuration with $u=u_0$.

The effective Langevin equation corresponding to (\ref{eq_dissip}) should
read
 \begin{equation}
 \label{eq_lan_per}
 \partial_t u(x,t)=-\epsilon u+D\partial_x^2 u + \partial_x^2 \eta(x,t)
 \end{equation} 

On large lengthscales noise averages out and we obtain the following 
dependence of the growth rate on the wave vector:
 \begin{equation}
 \label{eq_qfit} 
 \lambda(k)=-\epsilon-Dk^2,
 \end{equation}
 i.e. we should expect (for small $k$'s) the Lyapunov spectrum of the
perturbed system to be shifted downwards by $\delta\lambda=-\epsilon$ with
respect to the spectrum of the conserving system, while retaining the same
type of singularity. The total decay rate is then determined by a linear
combination (at least for small enough coupling) of the diffusion with
local dissipation. 

Comparing the spectra of Lyapunov exponents of the modified system (fig.
\ref{fig_lin}) and the original system (fig. \ref{fig_cube}(a)) for
$u=u_0=0.8$ we see that the numerically obtained spectra of the perturbed
system follow the prediction of the Langevin equation (\ref{eq_lan_per})
quite precisely for small ($\epsilon=0.01$) as well as for relatively
strong ($\epsilon=0.1$) perturbations. 

We expect the negative shift of the ``slow'' part of the spectrum to be
attributed to the dissipative nature of the perturbations used above. In
fact we could have a positive shift or no shift at all. In order to see
this we pick the nonlinear perturbation function $g(x)$ with the first
derivative which is not negative-definite: 
 \begin{equation}
 \label{eq_oscil}
 g(x)={\gamma^2(u_0-x)\over ((u_o-x)^2+\gamma^2)}
 \end{equation}
 with $\epsilon>0$. We used $\gamma^2=0.001$ and $0.01\le\epsilon\le 5$.

Equation (\ref{eq_1dmap}) does not hold anymore, but the numerical data
suggests that the dynamics of the system is ergodic and the averaged
density $u$ in the asymptotic state fluctuates about $u_0$ with
fluctuations being again of order few percent. 

Fig. \ref{fig_lor} shows the Lyapunov spectra of the perturbed system with
$u_0=0.8$. This figure suggests that the spectrum is indeed shifted in the
direction of positive rather than negative values of $\lambda$. A small
perturbation ($\epsilon=0.1$) results in a small distortion of the
original spectrum: the slope of the spectrum at the value $\lambda=0$ on
both positive and negative sides becomes nonzero, i.e. the singularity in
the density of Lyapunov exponents disappears. The slope increases with
$\epsilon$ and for sufficiently strong perturbation ($\epsilon\ge 2.0$)
all traces of the singularity vanish. 

These examples suggest that there are, in fact, two different aspects of
the singularity in the Lyapunov spectrum of a system with the conservation
law. The first one is the presence of a singularity at some value
$\lambda=\lambda_0$. The numerical data obtained suggests that the
singularity survives in the special case of the linear perturbation
function $g(x)$. A nonlinear perturbation results in the disappearance of
the singularity. The Jacobian of a perturbed system can be written as
 \begin{equation}
 \tilde{J}^n_{kk'}=J^n_{kk'}+\sum_{j\ge 1}{g^{(j)}(u_0)\over (j-1)!}
 \prod^{j-1}_{i=1}\sum_{k_i}u^n_{k_i}\delta(\sum^{j-1}_{l=1}k_l-k+k'),
 \end{equation}
 where $J^n_{kk'}$ is the Jacobian (\ref{eq_jacob}) of the conserving
system. If $g(x)$ is nonlinear the off-diagonal elements of
$\tilde{J}^n_{kk'}$ become of order $O(1)$ instead of $O(k^2)$. As a
result Fourier modes are no longer good as an approximation to the exact
eigenvectors even for small $k$. The wavevector $k$ can no longer label
the slow modes of the system and therefore there is no reason to expect
the singularity in the density of Lyapunov exponents to remain. It becomes
``smoothed out'' by the perturbation. 

The second aspect is the actual value of $\lambda_0$ in case a singularity
is present. We already saw that imposing linear perturbation
(\ref{eq_dissip}) made $\lambda_0$ become negative (cf. (\ref{eq_qfit})). 
This is a consequence of the shift of the ``slow'' part of the spectrum as
a whole in response to some local effects, e.g. dissipation.

We may therefore suggest, that since all known CML's with an additive
conserved quantity possess a spectrum of Lyapunov exponents distinguished
by the presence of the singularity at $\lambda=0$, a conservation law is a
sufficient condition for the existence of such a singularity. It seems to
be a necessary condition as well, at least in the class of models
(\ref{eq_cml}) studied in this paper. The singularity at $\lambda\ne0$
indicates that there is a mechanism of (local) dissipation concurrent with
diffusion and if this is eliminated the system becomes strictly
conserving. This refinement might be helpful when looking for hidden
conservation laws using Lyapunov spectra \cite{bohr}, numeric or
experimental. 

All three types of perturbation studied above are seen to have an effect
on the phase diagram. For $\epsilon=0$ phase diagrams in 3-dimensional
parameter spaces $(a,b,u)$ and $(a,b,u_0)$ obviously coincide. Numerical
data for all types of perturbation studied suggest that gradual increase
of the parameter $\epsilon$ makes the phase diagram of the perturbed
system change continuously. The boundaries of the phases shift, making
some of the phases shrink or completely disappear (e.g. sufficiently
strong perturbation of any type obliterates the phase L2). Other phases
may expand as the stability of their basic state is enhanced (as is the
case for the phase L1 and the uniform state in the presence of a
``dissipative'' perturbation). We can say that, although the phase diagram
is sensitive to the violations of the conservation law, it is robust with 
respect to sufficiently weak violations.

Since the changes in the phase diagram provoked by perturbations are found
to be continuous, it is reasonable to expect that the phase transitions
which were 1-order in the conserving system will remain 1-order after a
perturbation is imposed. This leaves us an opportunity to observe how the
characteristic exponents change with increasing perturbation. In our model
the continuous phase transition at the boundary T2-L1 seems to be a
promising point of investigation. Numerical data we have at the moment
does not allow us to answer an important question of whether the
universality classes change if the conservation law is violated, but we
plan to return and investigate this later in more detail. 

\section{Conclusions}
\label{sec_conclusion}

We have systematically investigated the properties of a coupled map
lattice with dynamics constructed to satisfy a conservation law and to
show spatiotemporal chaos. 

The conserved quantity provides an additional control parameter: as its
value is changed, a rich phase diagram with a number of phase transitions
between ordered and disordered states is found. Both continuous and
discontinuous transitions occur, as in coupled map lattices without a
conservation law. The basic structure of the phase diagram is given by the
linear stability boundaries of the ordered phases, although near the
discontinuous transitions bistability may occur. Increasing the
nonlinearity, determined by the parameter $b$ (see fig.1(b)), renders the
spatially uniform ordered phase (phase $L1$) unstable. For $0<u<1/2$ the
linear instability of the uniform state occurs via spatial period doubling
(the zone boundary mode goes unstable). For $a<u<1/2$ the transition is
immediately to a chaotic state, which takes the form of an increasing
number of ``turbulent'' regions of roughly fixed size (defects) moving
through the ``laminar'' background. This transition is continuous with a
diverging correlation length and other scaling approaching the ordered
state. For $0<u<a$ the transition to chaos is through a series of two
subharmonic bifurcations, passing first to an intermediate 2-cycle state
(the $L2$ phase). The onset of chaos from $L2$ is hysteretic. A complete
subharmonic cascade is not observed. For $1/2<u<1$ the instability of the
$L1$ phase occurs through modes of all wave vectors going unstable
together, and the appearance of chaos is also hysteretic with frozen
chaotic domains developing. 

As in thermodynamic systems, the phase transitions can be conveniently
described by the use of order parameters, although since we are concerned
with the growth of {\em disorder} the choice of the appropriate order
parameter here is by no means obvious. In the case of the continuous
transition we find that a number of proposed order parameters scale with
distance to the transition point, thus allowing the evaluation of critical
exponents, which may help to pin down whether {\em universality classes}
for the onset of spatiotemporal chaos exist. An interesting question is
whether the conservation law, which clearly affects the dynamic
correlations, also changes the exponents and the universality class (if
such a classification exists) of the phase transitions. The transition
from $L1$ to $T2$ in the conserving model and in perturbed versions where
the conservation law is weakly violated should provide a good arena for
investigating this. 

A symbolic description of the dynamics, reducing the complex states to
regions of laminar (ordered) and turbulent (fluctuating) regions is useful
in describing the chaotic states near the transitions. In particular we
see that the onset of chaos always happens in the form of turbulent
regions gradually spreading over the laminar background. Whenever the
chaos appears to grow continuously, the turbulent fluctuations appear in
the form of turbulent defects, usually propagating across the system, with
size $l_t$ of order a few lattice spacings. Similar results were obtained
by Kaneko \cite{kaneko} for the non-conserving CML, although in our case
there seems to be a stronger tendency for the defects to propagate with
constant velocity. 

We have also studied the Lyapunov eigenvalues and eigenvectors of the
chaotic states. A conspicuous feature of conserving models is the
singularity in the density of Lyapunov exponents around $\lambda =0$.
There is growing evidence that the singularity is associated with the
existence of Lyapunov eigenvectors that are labelled by the wave vector
$\vec{k}$ for small $k$ --- the singularity is then the usual Van Hove
singularity coming from mode counting and the assumption of a smooth
spectrum at small $k$, i.e. $\lambda =\lambda _{0}-D_{eff}k^{2}$. This
leads immediately to a $|\lambda -\lambda _{0}|^{-1/2}$ singularity in the
density of exponents. 

Some evidence for this idea comes from the work of Bohr et al. \cite{bohr}
who studied some models where the Lyapunov spectrum can be exactly
calculated, and in which the eigenvectors are trivially Fourier modes,
since the Jacobian is independent of the dynamic variables. In these
models the singularity can occur at nonzero $\lambda _{0}$, and the
singularity may be to larger or smaller $\lambda $. 

For the conserving (and nontrivial) maps we have studied, the labelling of
the Lyapunov eigenvectors by the wave vector and the expression for the
Lyapunov spectrum, with $\lambda _{0}=0$ and $D$ positive, is suggested by
a hydrodynamic analysis, which associates the Lyapunov eigenvectors with
the diffusively decaying modes given by a long wavelength Langevin
description. The Fourier power spectrum of the Lyapunov eigenvectors
provides support for this explanation: for small negative $\lambda $ the
power spectrum corresponding to the m-th eigenvalue $\lambda _{m}$ is
indeed strongly peaked around a small $k$ with $k\propto (m-m_{0})2\pi /L$
with $\lambda _{m_{0}}=0$. Note however that the diffusion constant
estimated form the Lyapunov spectrum may be significantly different from
other estimates, and the proportionality constant $\alpha $ is greater
than $0.5$ which would be the value from simply counting the Fourier
modes. Indeed, there is also a concentration of spectral power towards
small $k$ on the positive $\lambda $ side, with a peak wave vector which
again appears to scale linearly with $m_{0}-m$ for $m$ sufficiently close
to $m_{0}$. Hence the long wavelength Fourier modes contribute appreciably
to the Lyapunov vectors corresponding to small positive as well as
negative exponents, which explains the deviation of $\alpha $ from $0.5$.
There is no understanding from the hydrodynamic approach of the positive
eigenvalue long-wavelength modes and associated singularity in the density
of exponents. 

Adding the linear perturbation (\ref{eq_dissip}) makes $\lambda _{0}$
nonzero, but $\vec{k}$ remains a good label. Nonlinear perturbations that
eliminate the conservation law however destroy the singularity, and
presumably in this case $\vec{k}$ is no longer a good label. Thus the
conservation law appears to be a sufficient condition for Lyapunov
spectrum singularity, but a complete quantitative understanding of this
association remains lacking. 

\acknowledgements

R.O.G. would like to thank Professor N. R. Corngold for many helpful 
discussions. The authors gratefully acknowledge the support of the 
National Science Foundation under Grant No. DMR9013984.


\vfill\eject

\vbox{ 
\begin{figure} 
\centering 
\mbox{
\psfig{figure=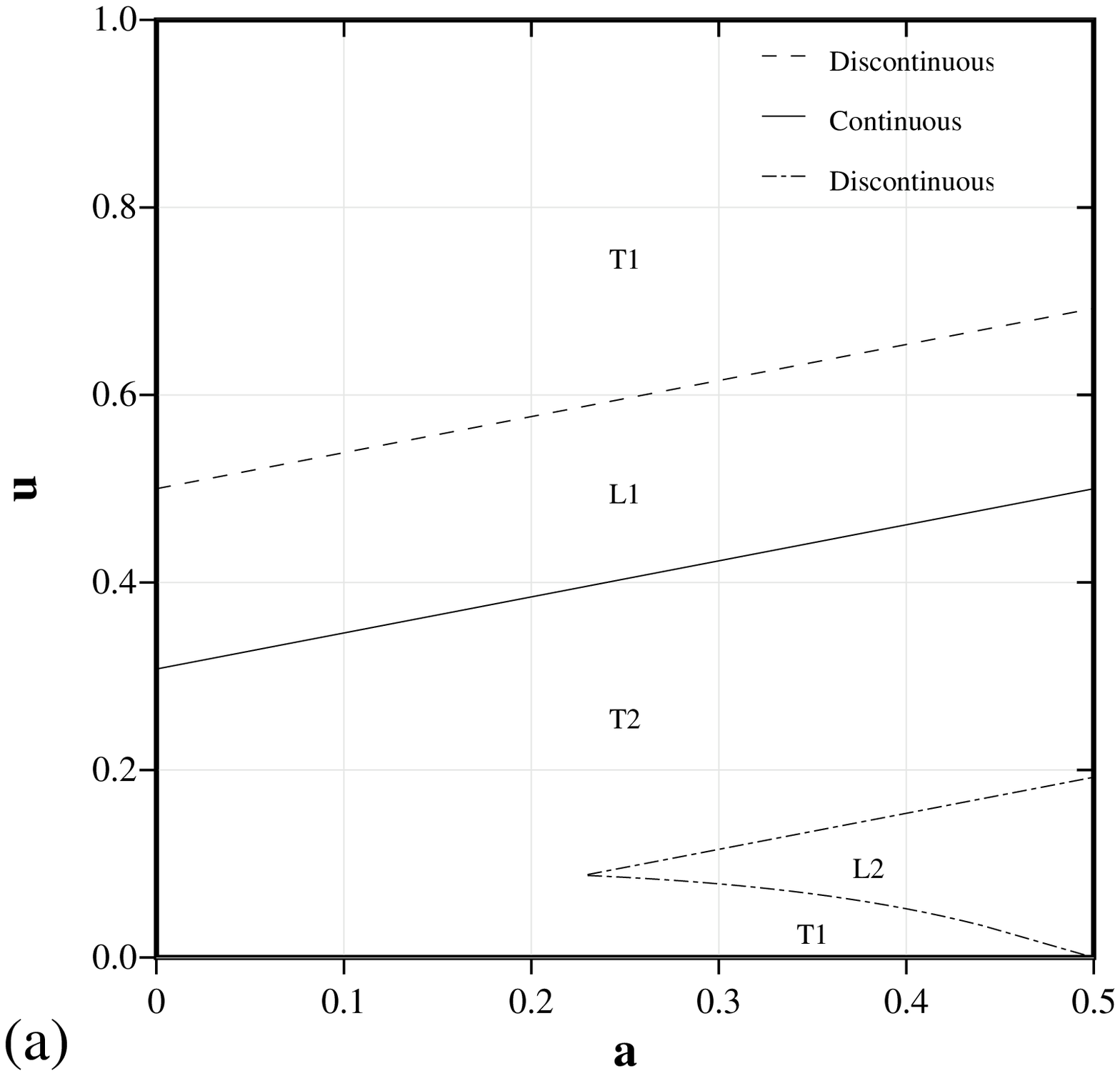,width=3in}\qquad 
\psfig{figure=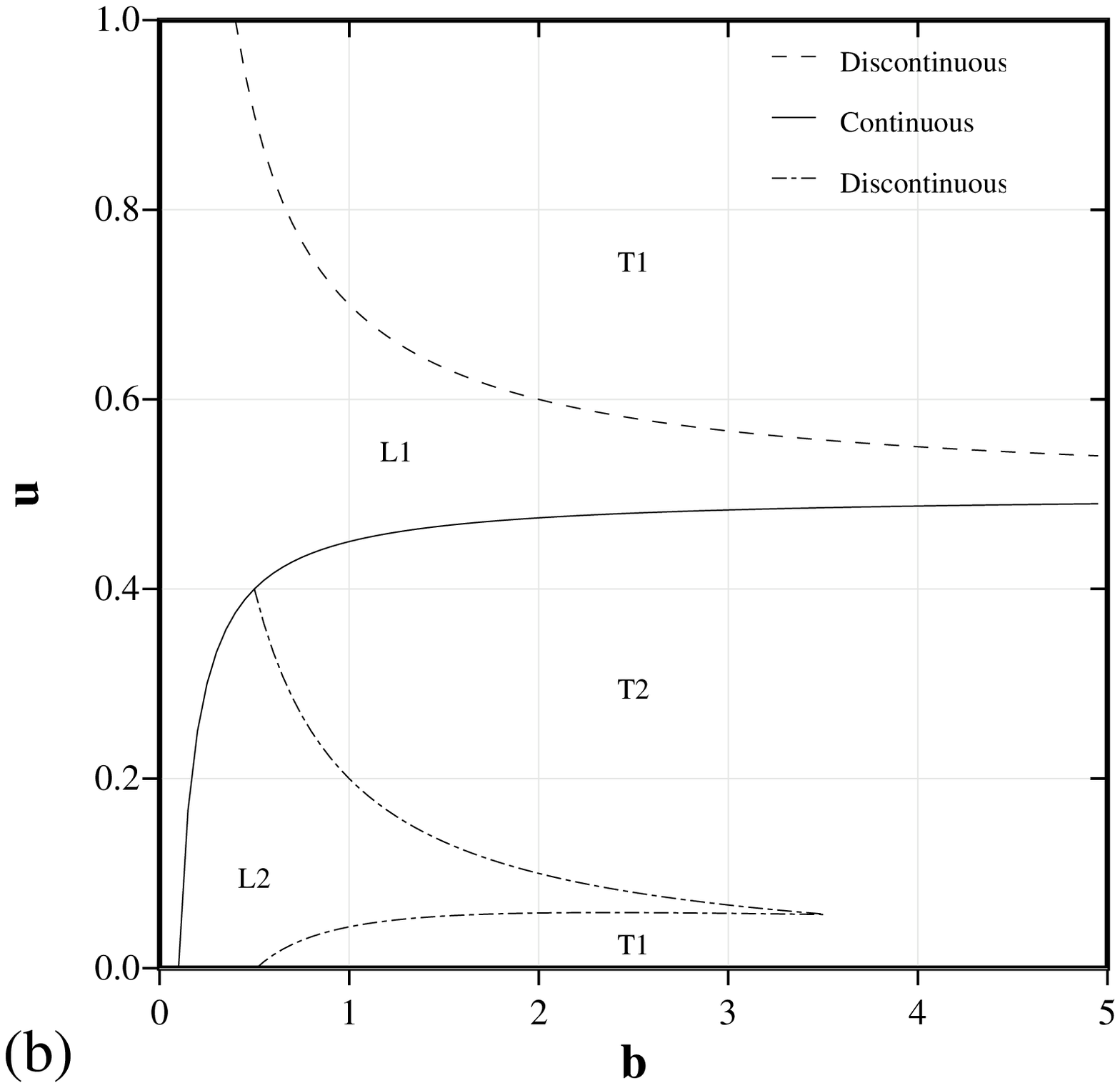,width=3in}}
\vskip 3mm
 \caption{Cross-sections of the phase diagram in a 3-dimensional parameter
space (a) $u$ vs. $a$ at $b=1.3$; (b) $u$ vs. $b$ at $a=0.4$. The solid
line corresponds to a continuous phase transition. The dashed and the
dotted lines denote the phase boundaries where discontinuous phase
transitions between chaotic and non-chaotic states occur.}
 \label{fig_phase}
\end{figure}}

\vbox{
\begin{figure}
\centering
\mbox{
\psfig{figure=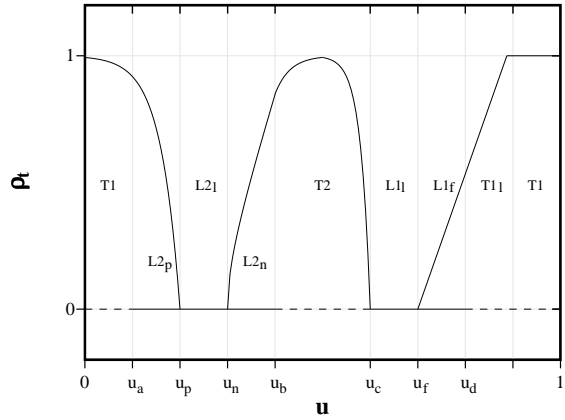,width=3in}}
\vskip 3mm
\caption{
  1-dimensional cross-section of the phase diagram. Measure of the set of
turbulent domains, $\rho_t$, is plotted schematically (actual numeric
results are presented in fig. \ref{fig_dimension}) as a function of
parameter $u$. Solid lines correspond to the values for the linearly
stable asymptotic states and dashed lines correspond to the unstable
asymptotic states. Two different linearly stable states coexist in the
sub-phases $L2_p,\ L2_n,\ L1_f$. A locked chaotic state always forms in
$T1_l$ and for some initial states in $L1_f$ as well. }
 \label{fig_hyst}
\end{figure}}

\vbox{
\begin{figure}
\centering
\mbox{
\psfig{figure=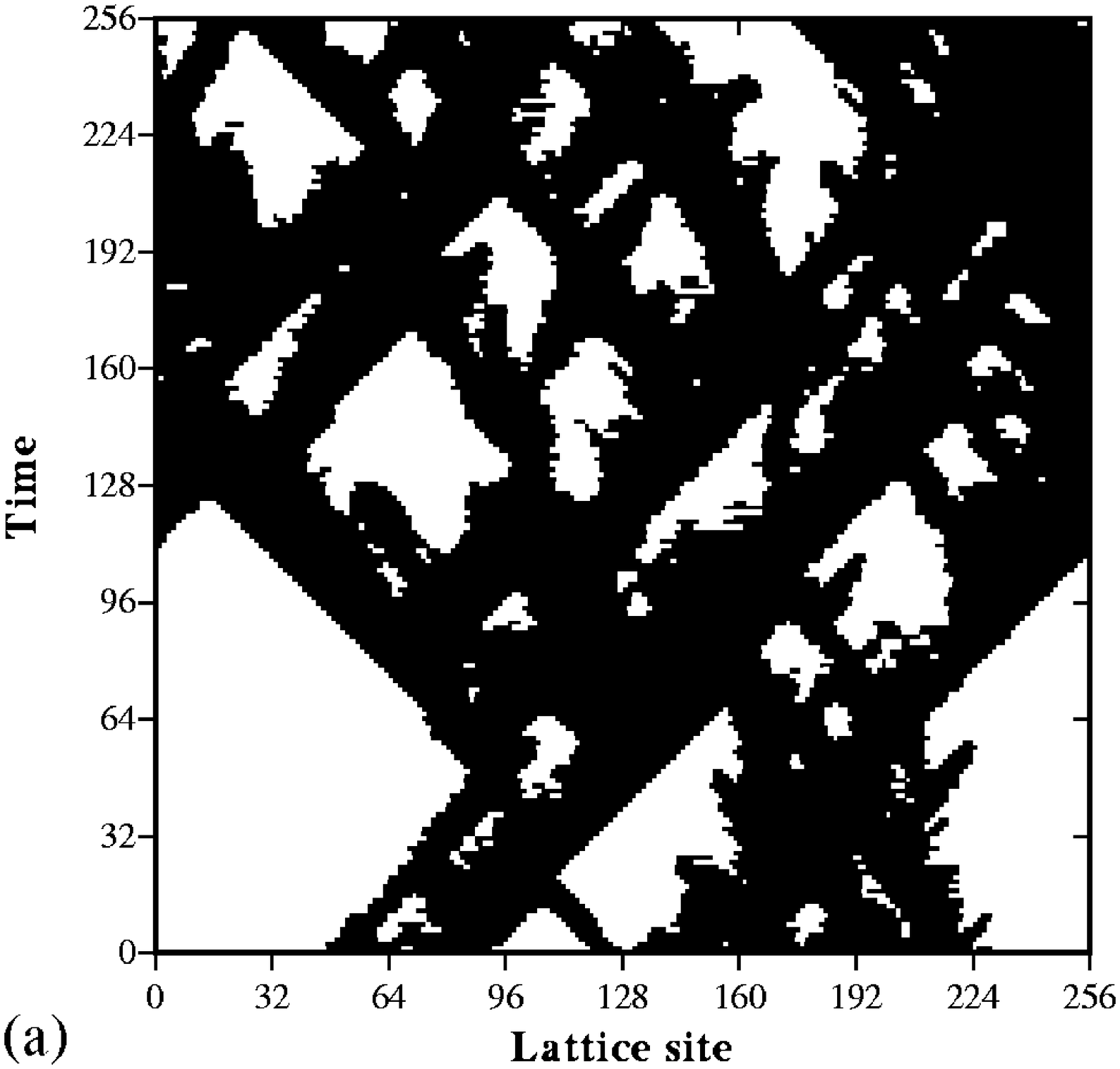,width=3in}\qquad
\psfig{figure=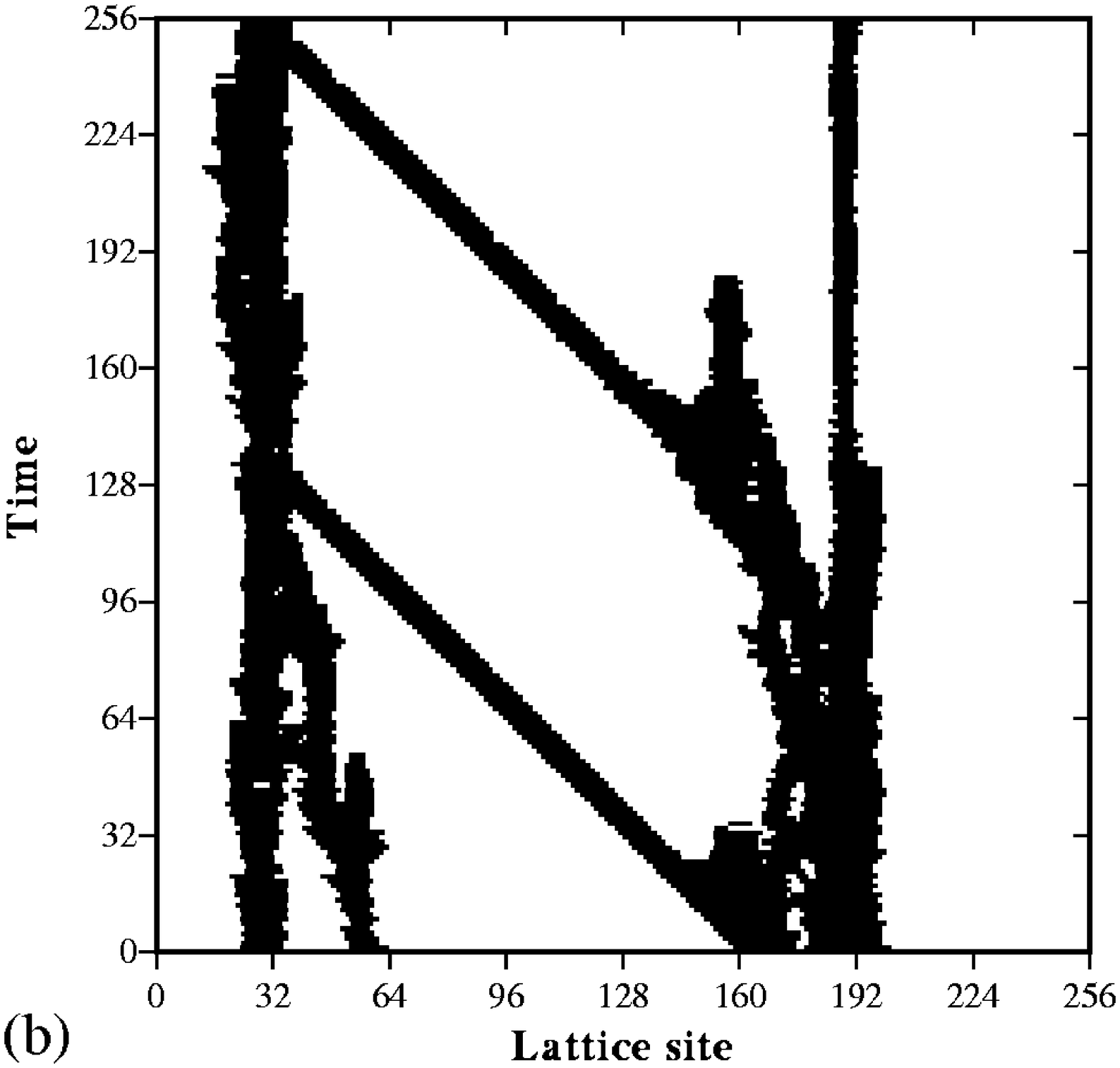,width=3in}}
\vskip 3mm
\centering
\mbox{
\psfig{figure=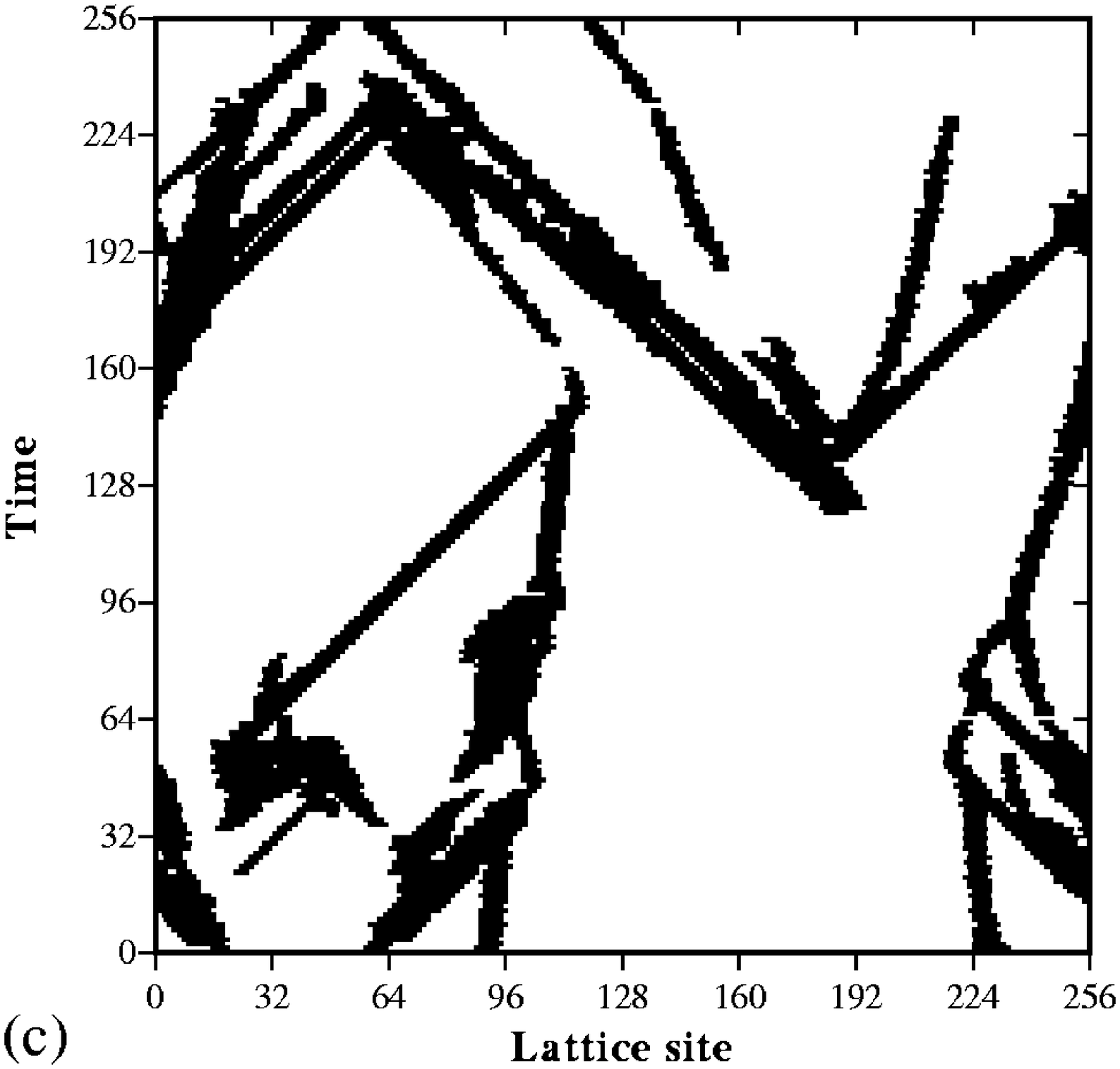,width=3in}\qquad
\psfig{figure=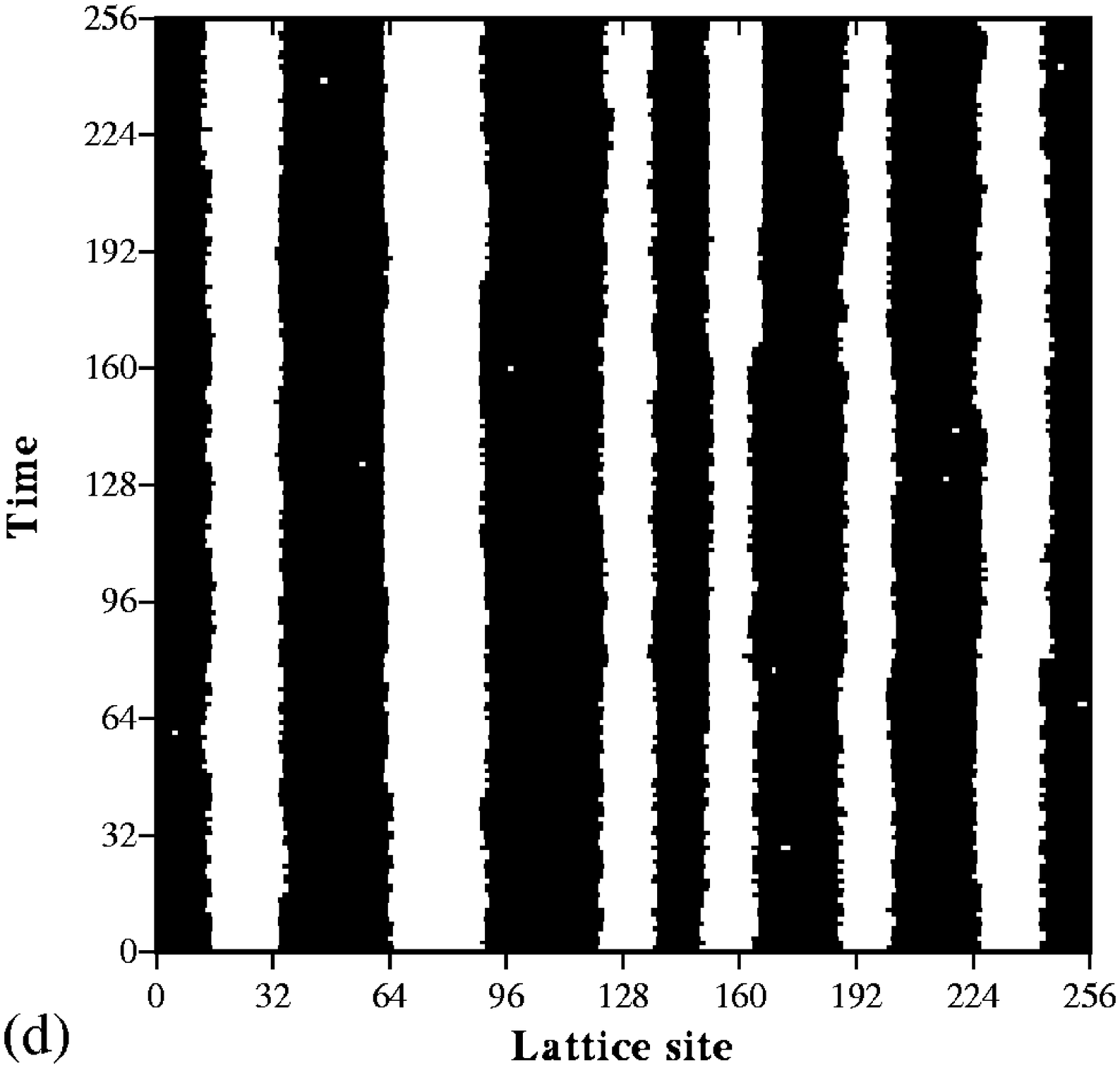,width=3in}}
\vskip 3mm
 \caption{Symbolic representation of the system dynamics close to the
phase transition points: turbulent sites are marked black and laminar
sites are white. (a) ``percolating'' state at $u=0.06$; (b) ``nuclear''
state at $u=0.09$; (c) defect dominated state at $u=0.46$; (d) frozen
pattern at $u=0.66$. Lattice size is 256.} \label{fig_pattern}
\end{figure}}

\vbox{
\begin{figure}
\centering
\mbox{
\psfig{figure=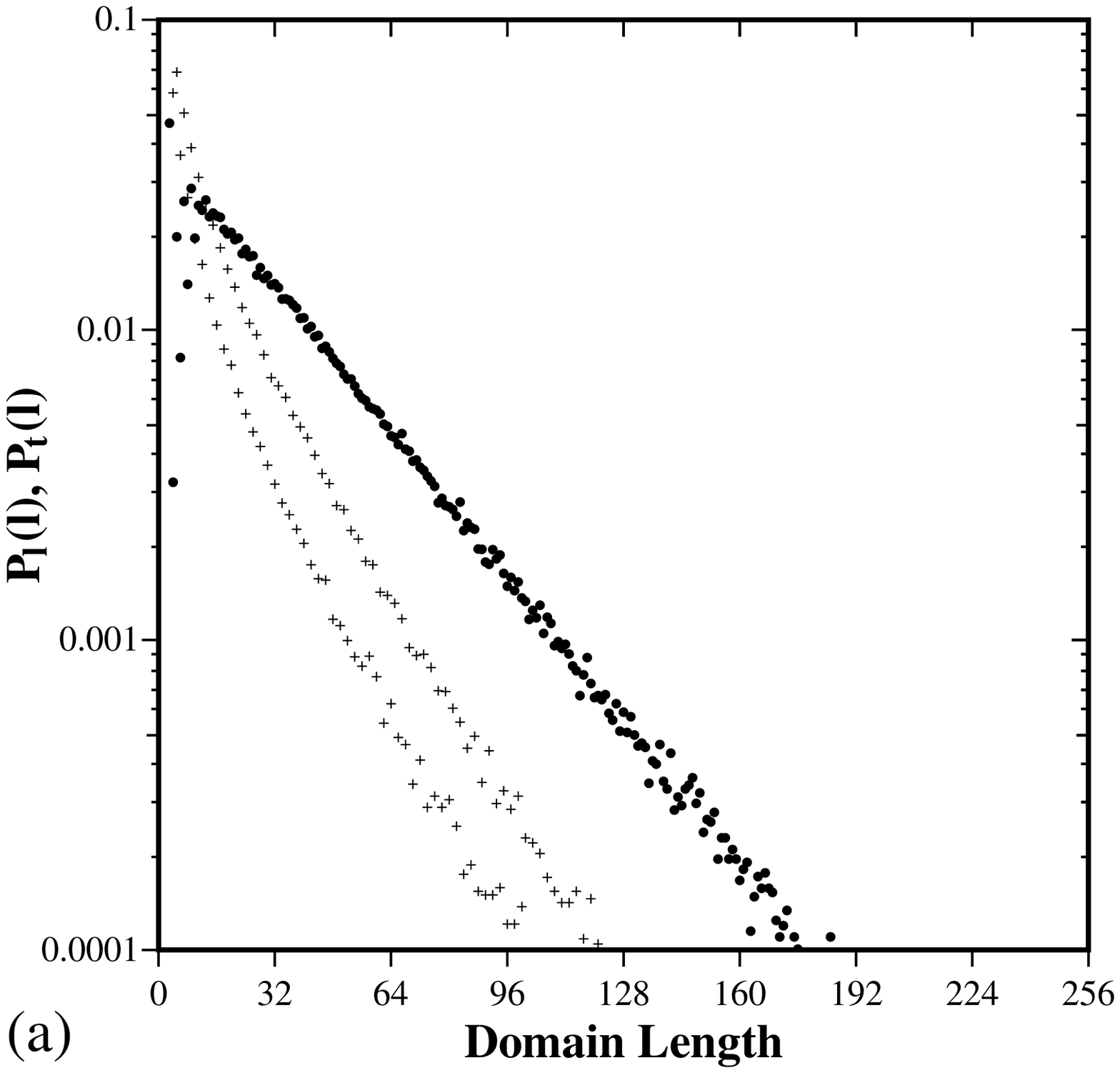,width=3in}\qquad
\psfig{figure=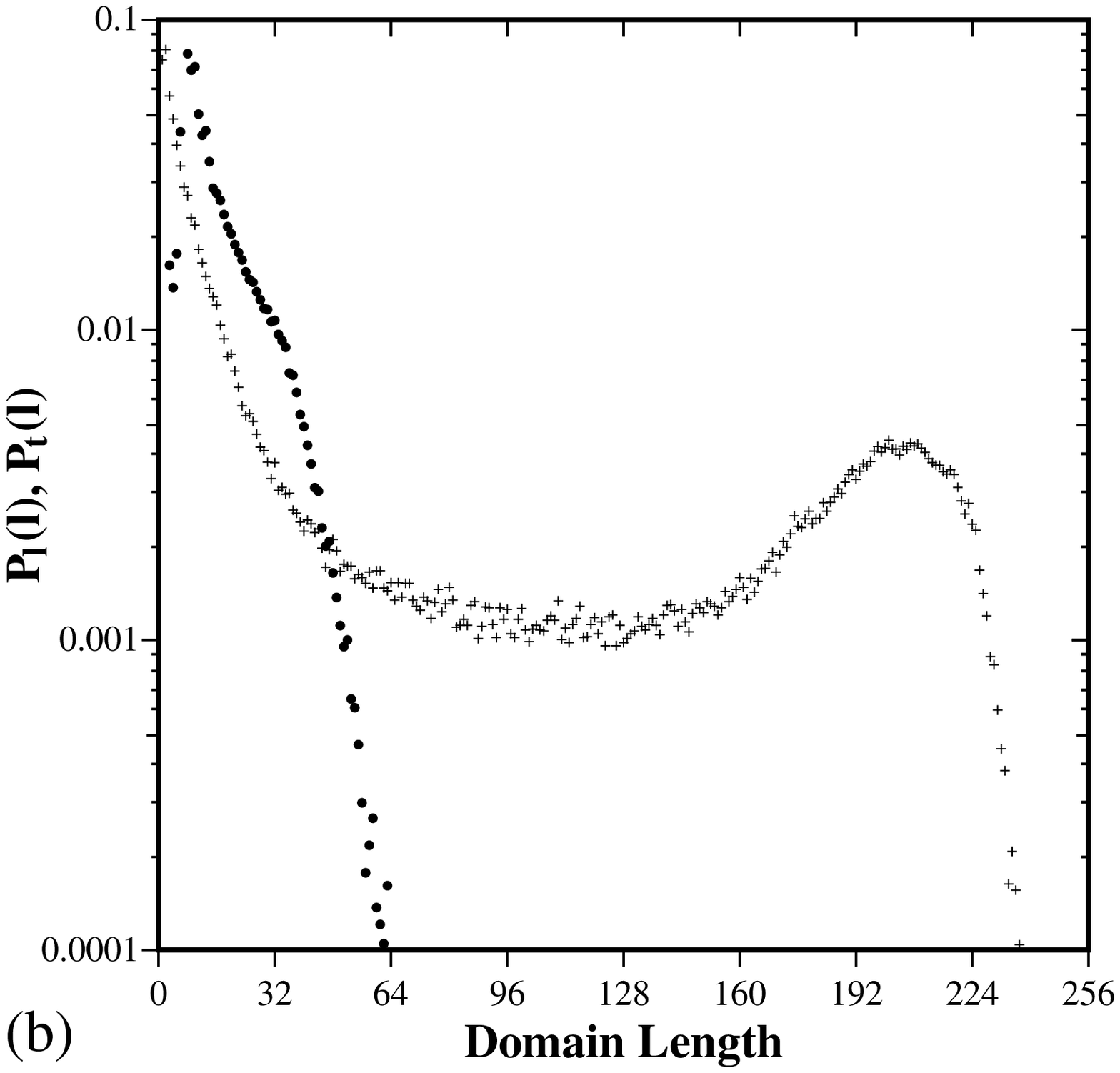,width=3in}}
\vskip 3mm
\centering
\mbox{
\psfig{figure=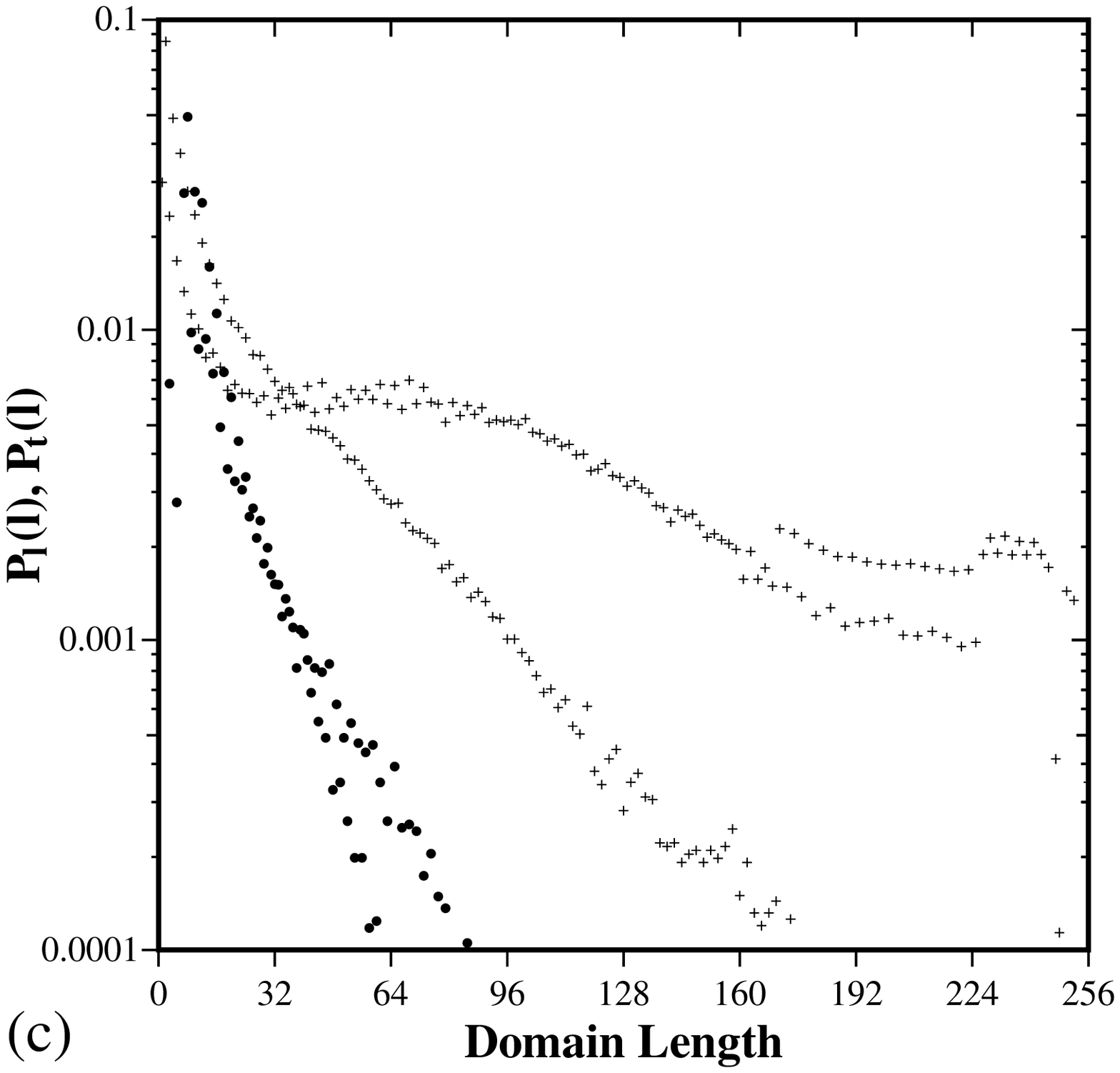,width=3in}\qquad
\psfig{figure=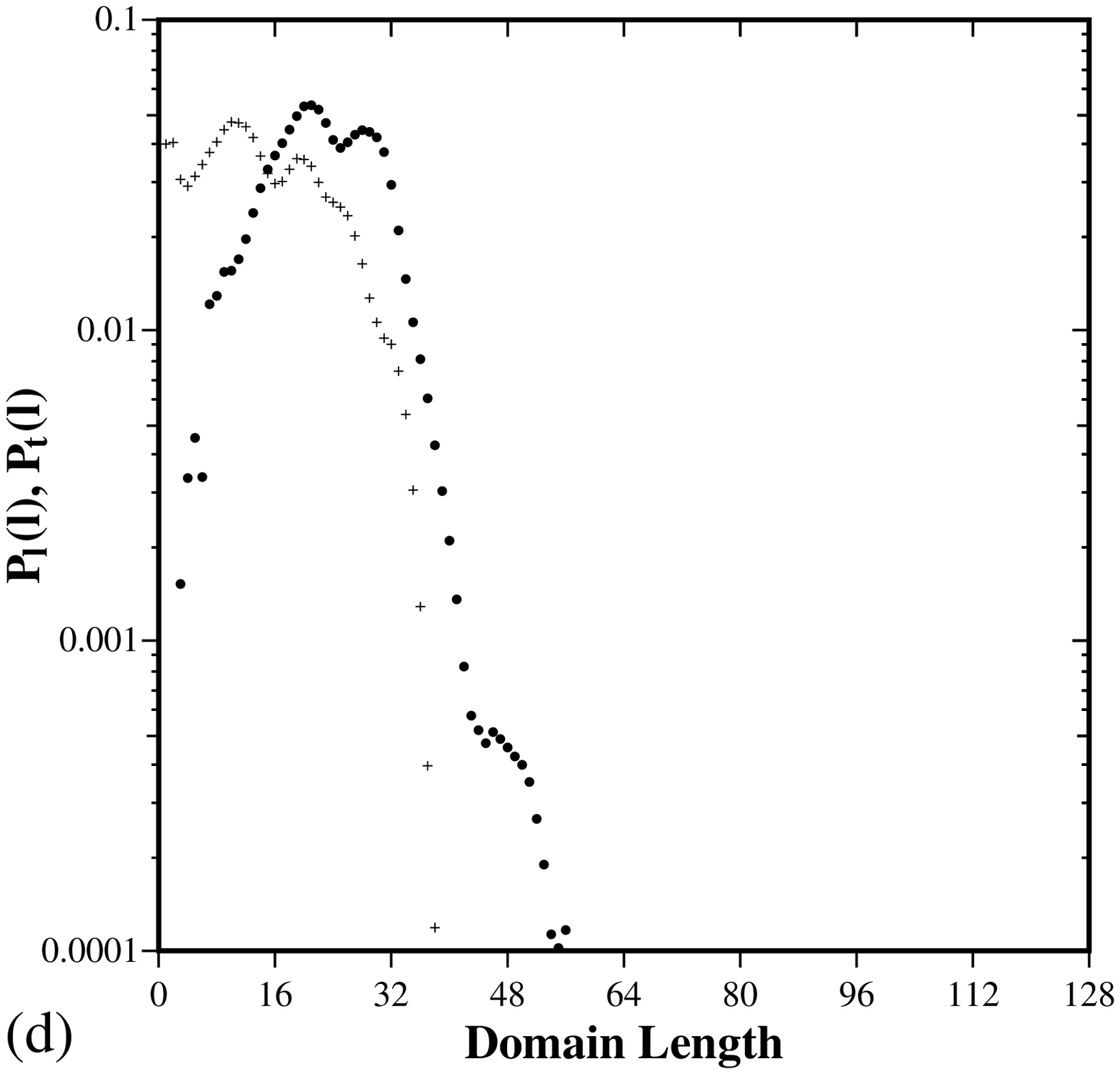,width=3in}}
\vskip 3mm
 \caption{Domain length probability distribution ({\scriptsize +} -
laminar $P_l(l)$, $\bullet$ - turbulent $P_t(l)$) in 
(a) ``percolating'' state at $u=0.06$,
(b) ``nuclear'' state at $u=0.09$,
(c) defect dominated state at $u=0.46$,
(d) frozen pattern at $u=0.66$.
Lattice size is 256.}
 \label{fig_domain}
\end{figure}}

\vbox{
\begin{figure}
\centering
\mbox{
\psfig{figure=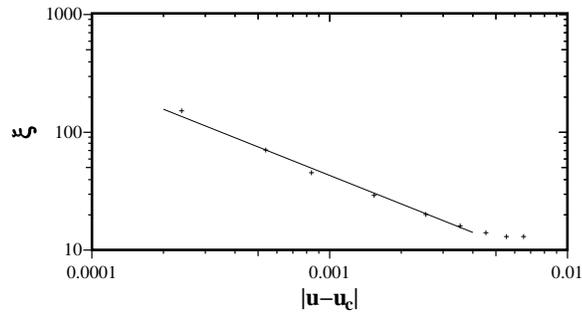,width=3in}}
\vskip 3mm
\caption{Correlation length diverges near the continuous phase transition}
\label{fig_corl}
\end{figure}}

\vbox{
\begin{figure}
\centering
\mbox{
\psfig{figure=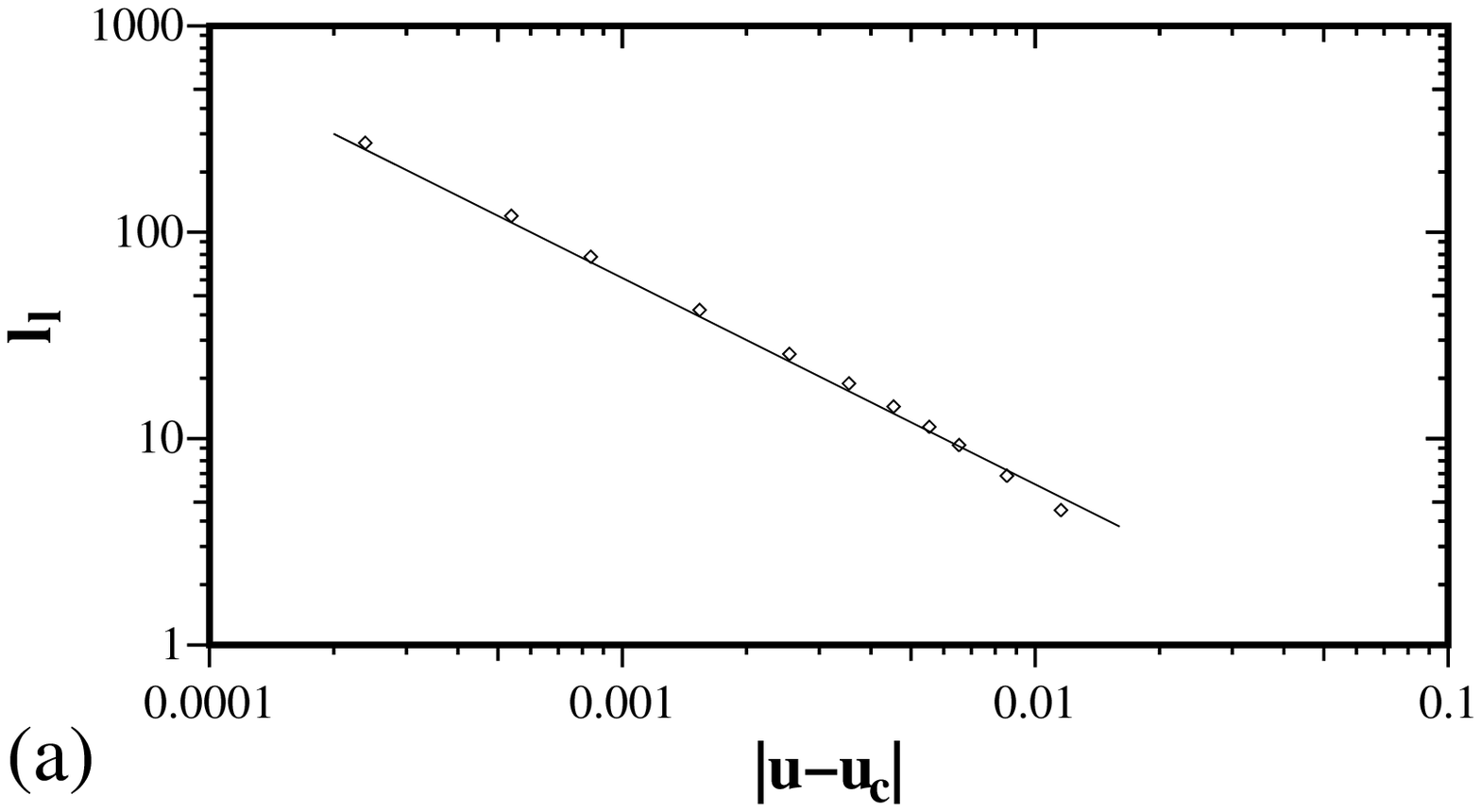,width=3in}\qquad
\psfig{figure=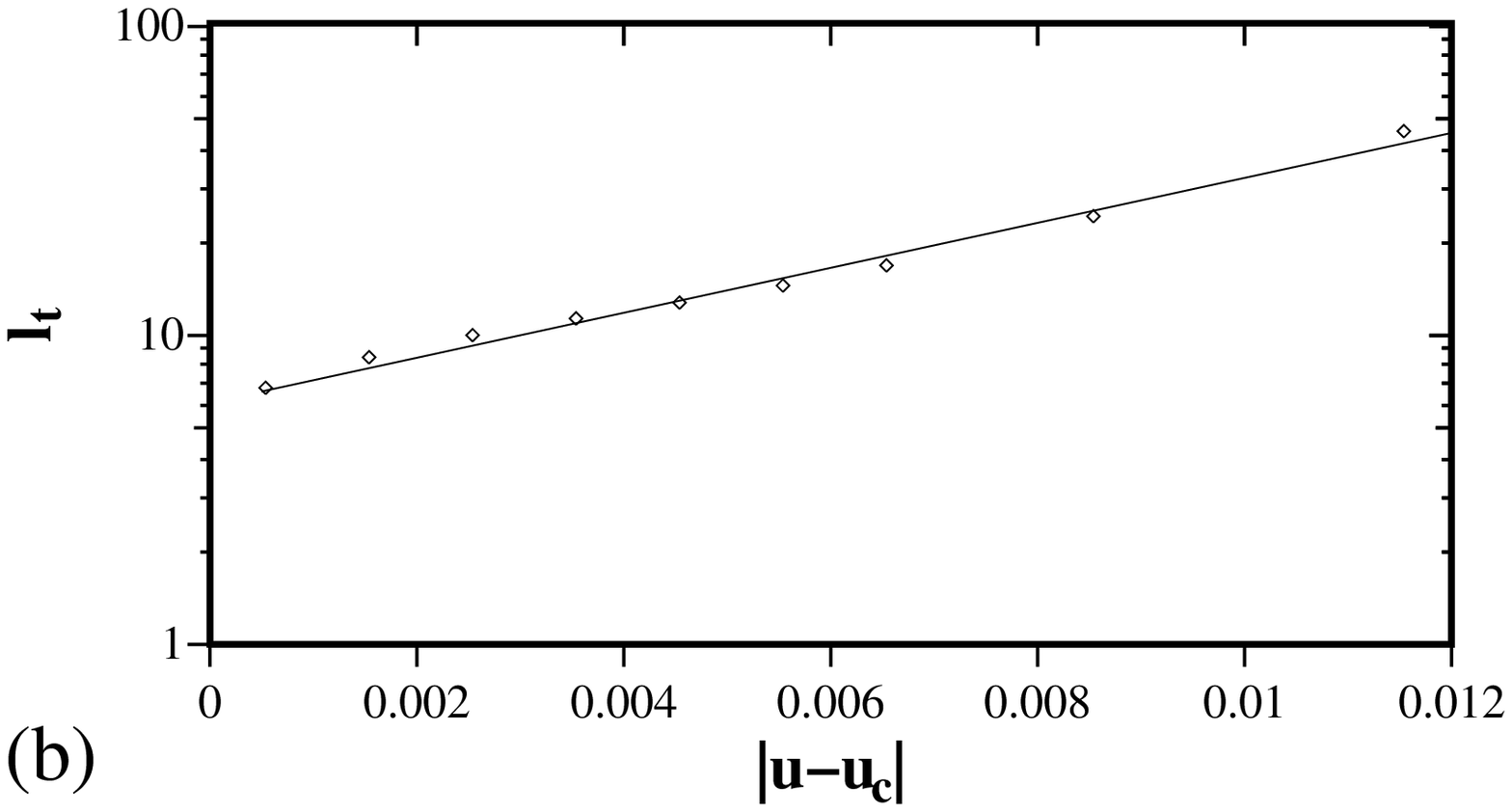,width=3in}}
\vskip 3mm
 \caption{Critical behavior of average domain lengths:
(a) length of laminar domains diverges algebraically,
(b) length of turbulent domains converges exponentially} 
\label{fig_length} 
\end{figure}}

\vbox{
\begin{figure}
\centering
\mbox{
\psfig{figure=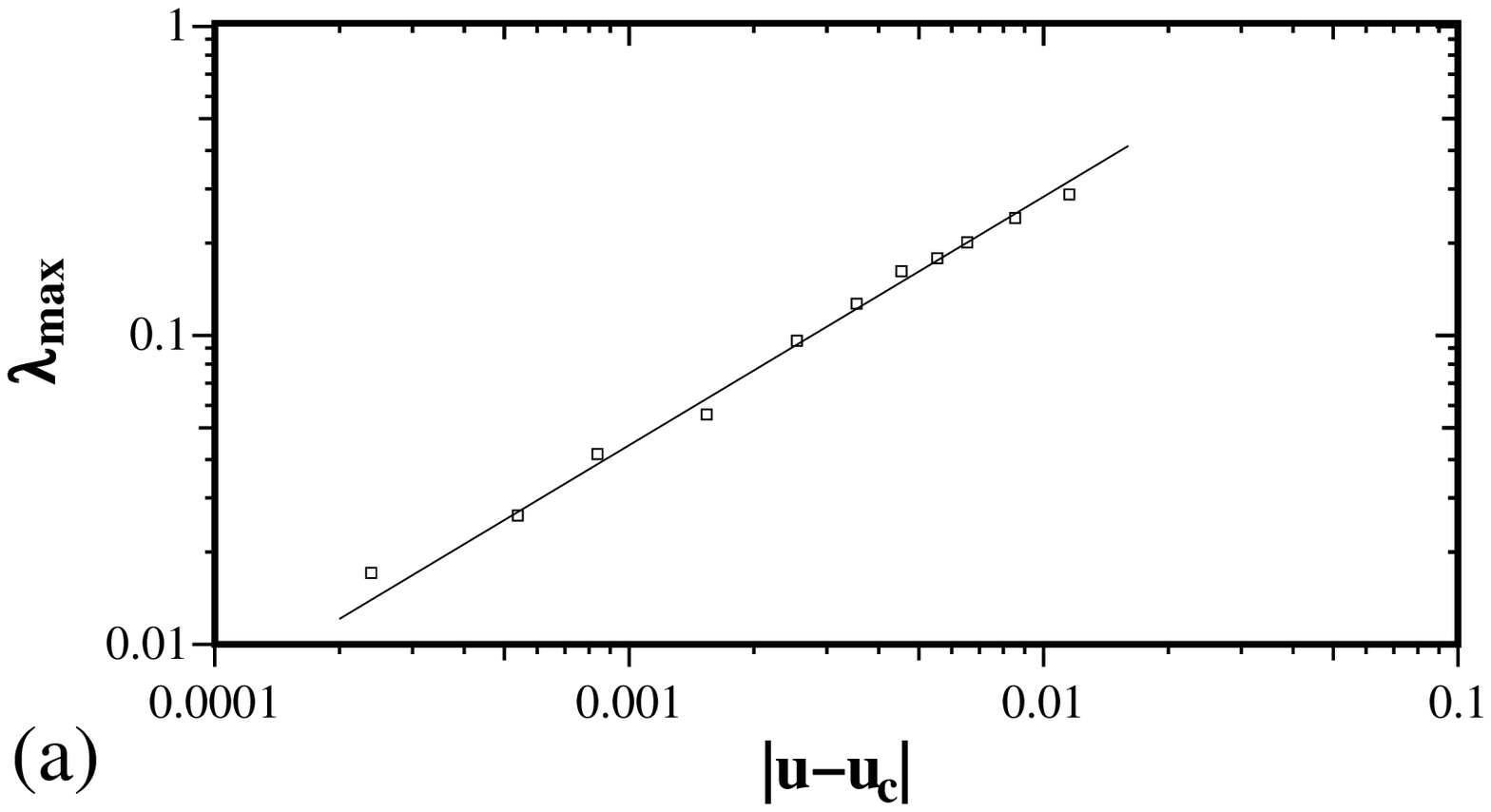,width=3in}\qquad
\psfig{figure=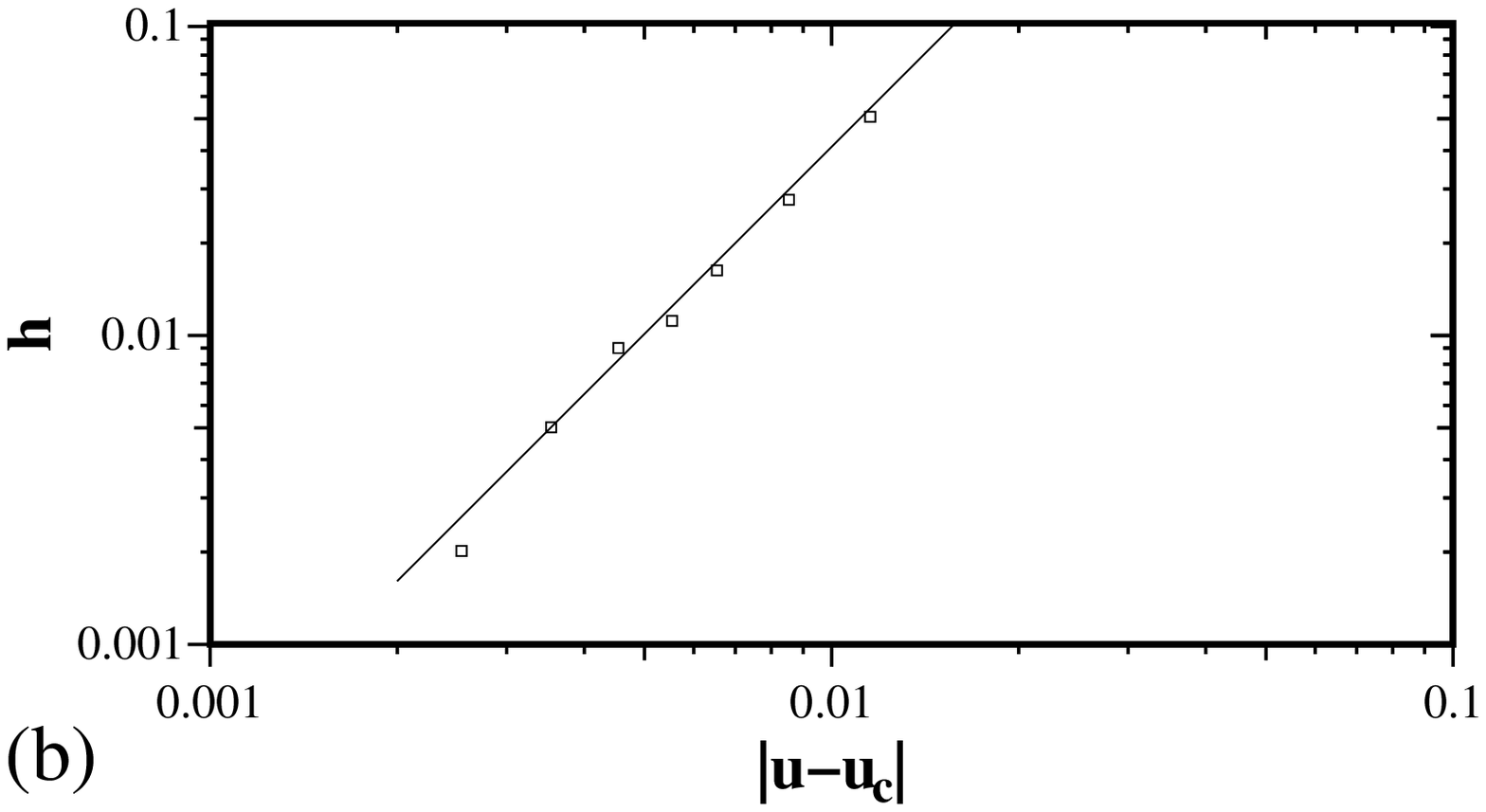,width=3in}}
\vskip 3mm
\centering
\mbox{
\psfig{figure=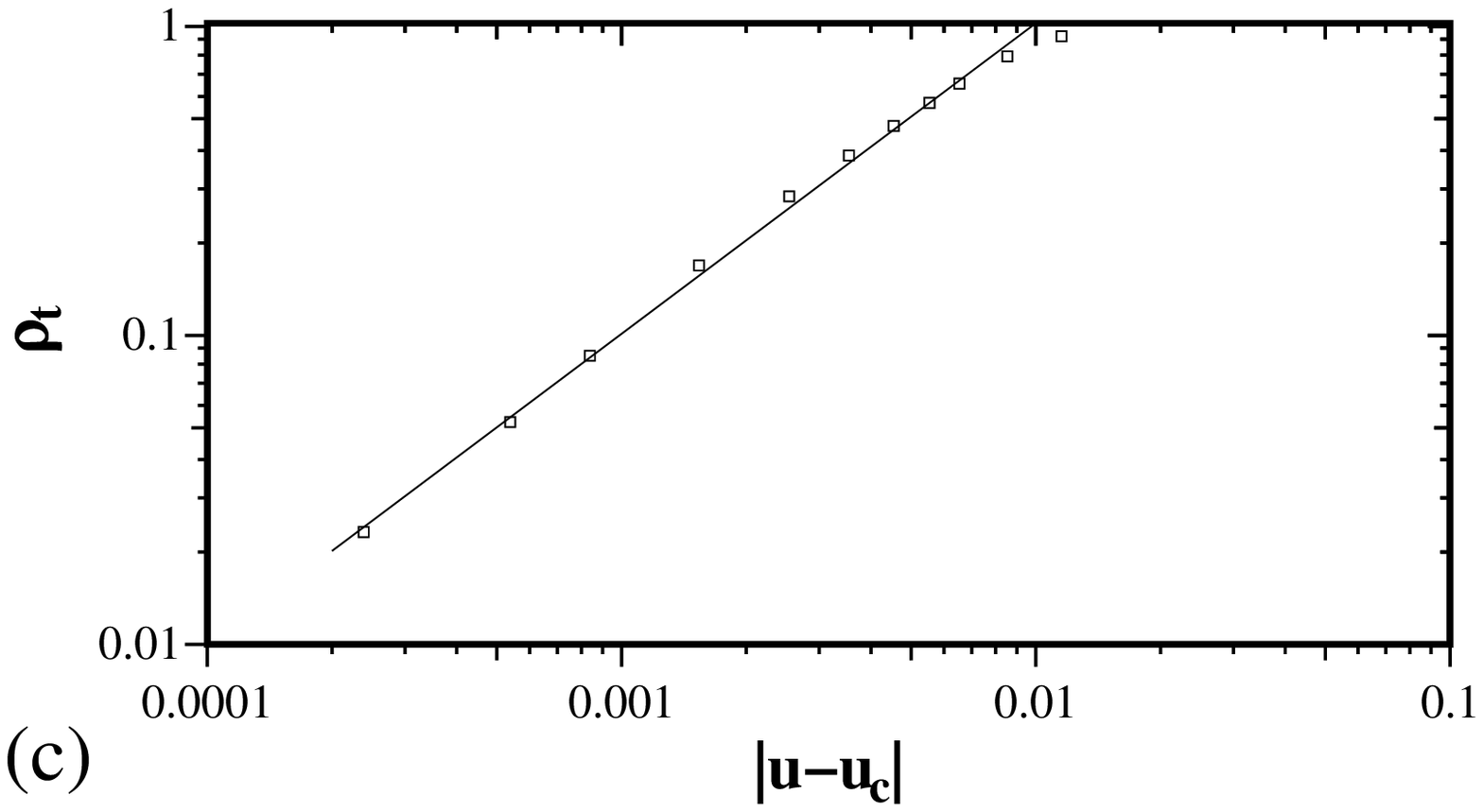,width=3in}\qquad
\psfig{figure=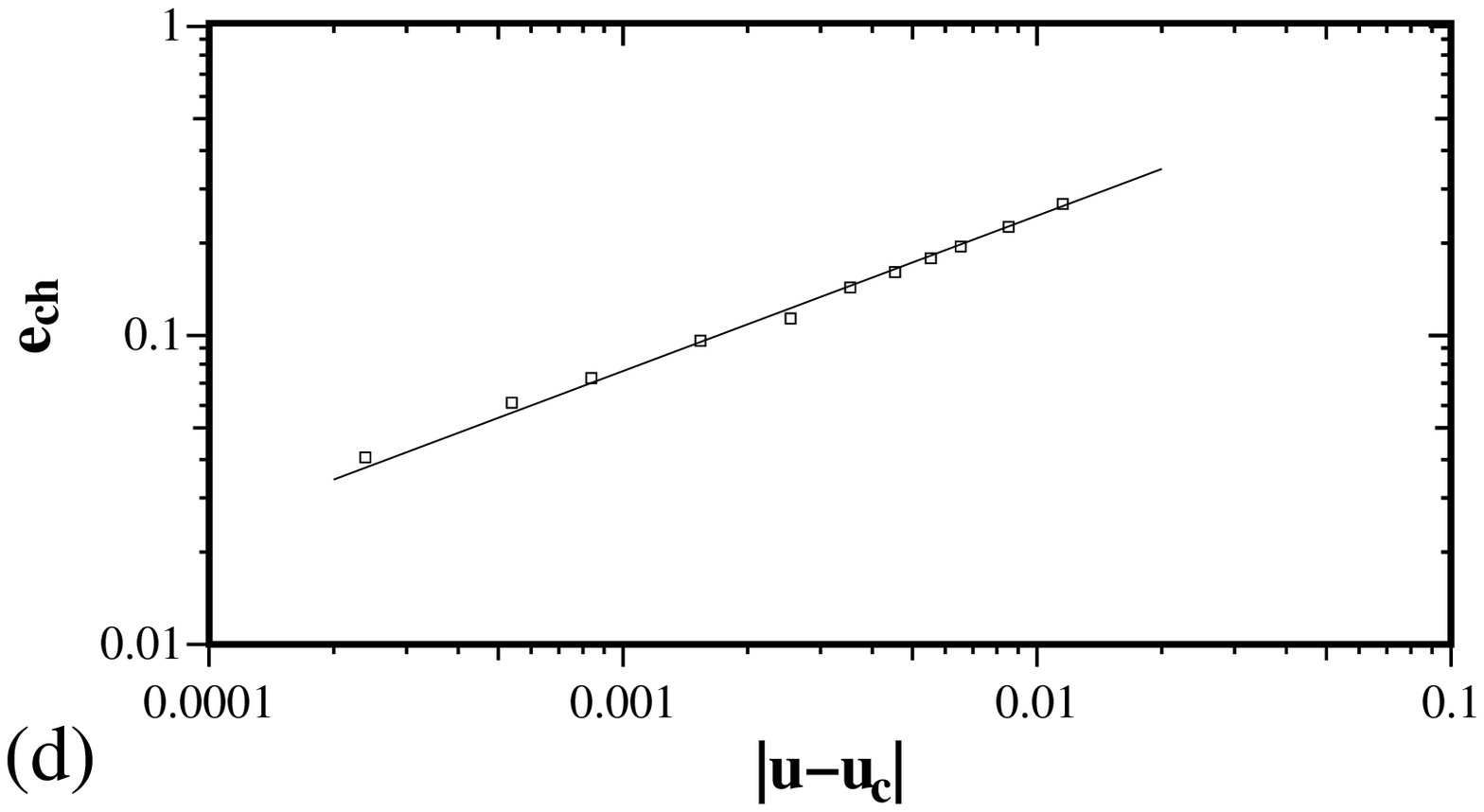,width=3in}}
\vskip 3mm
 \caption{Scaling of order parameters at the continuous phase transition 
calculated on the lattice with 2048 sites: 
(a) maximal Lyapunov exponent $\lambda_{max}$,
(b) Kolmogorov-Sinai entropy density $h$,
(c) measure of the turbulent set $\rho_t$,
(d) intensity of chaotic modes $e_{ch}$} 
\label{fig_oparam} 
\end{figure}}

\vbox{
\begin{figure}
\centering
\mbox{
\psfig{figure=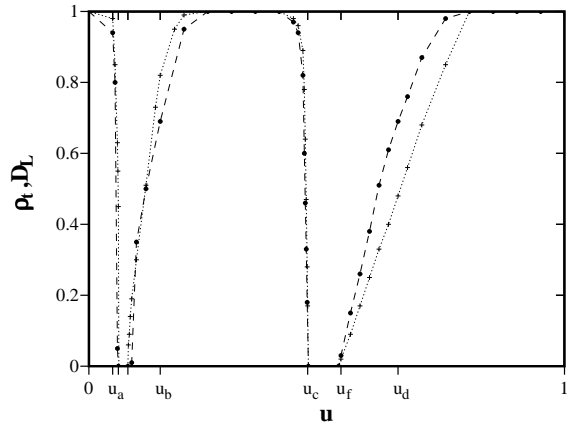,width=3in}}
\vskip 3mm
 \caption{Lyapunov dimension $D_L$ ($\bullet$) and turbulent set measure
$\rho_t$ ({\scriptsize +}) as functions of the conserved quantity $u$, 
calculated on the lattices with 128 and 2048 sites respectively. Only the 
values in the chaotic state are shown if two attractors coexist.}
 \label{fig_dimension} 
\end{figure}}

\vbox{
\begin{figure}
\centering
\mbox{
\psfig{figure=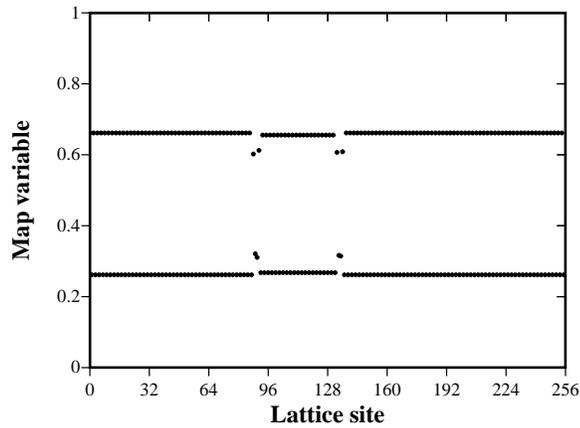,width=3in}}
\vskip 3mm
 \caption{Typical lattice configuration near the continuous phase
transition ($u=0.46$): two large laminar domains with different amplitudes
$A_1$ and $A_2$ are separated by two turbulent defects of the same width.
Lattice size is 256.}
 \label{fig_lattice} 
\end{figure}}

\vbox{ 
\begin{figure} 
\centering 
\mbox{
\psfig{figure=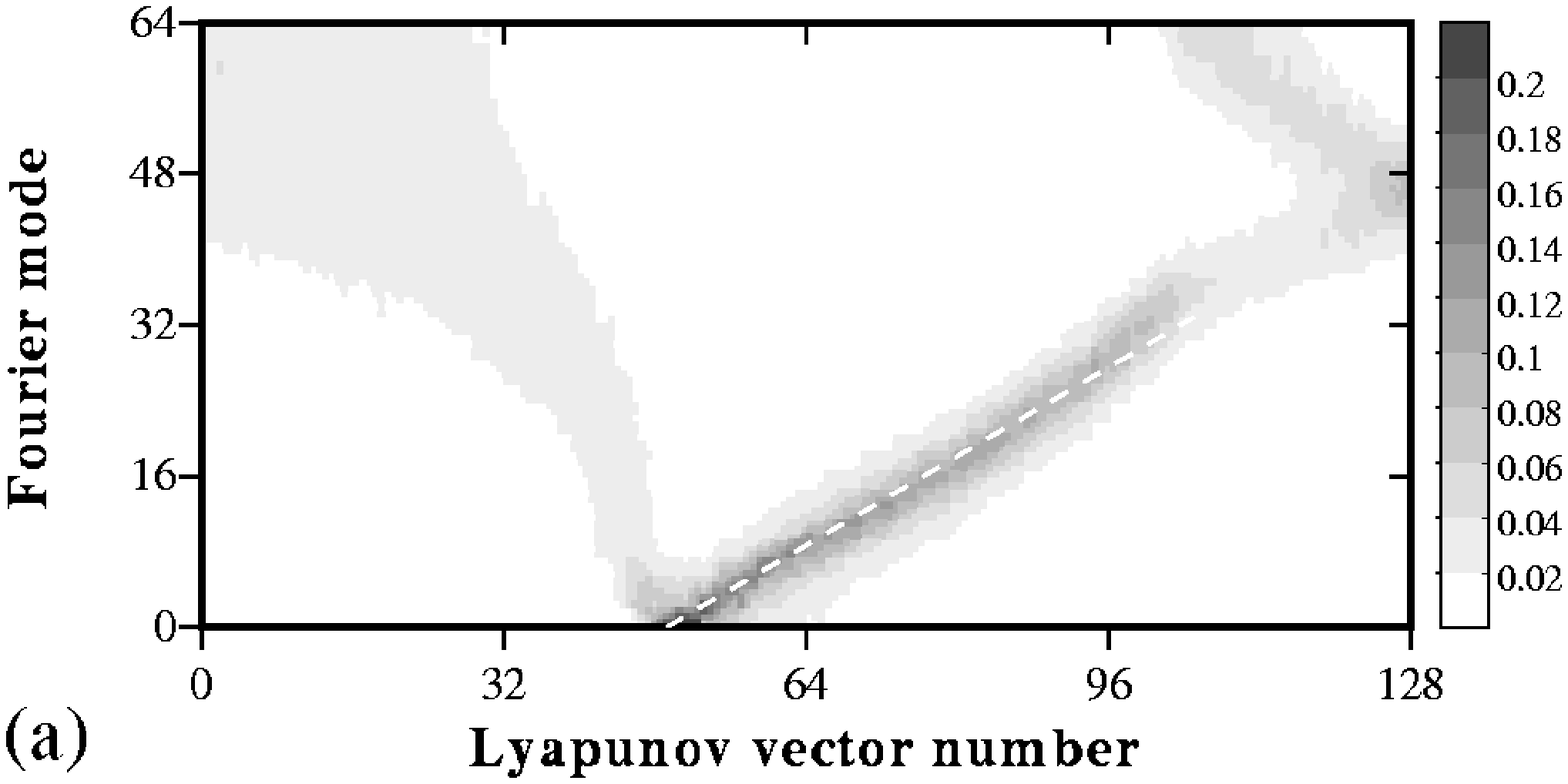,width=3in}}
\vskip 3mm
\centering
\mbox{
\psfig{figure=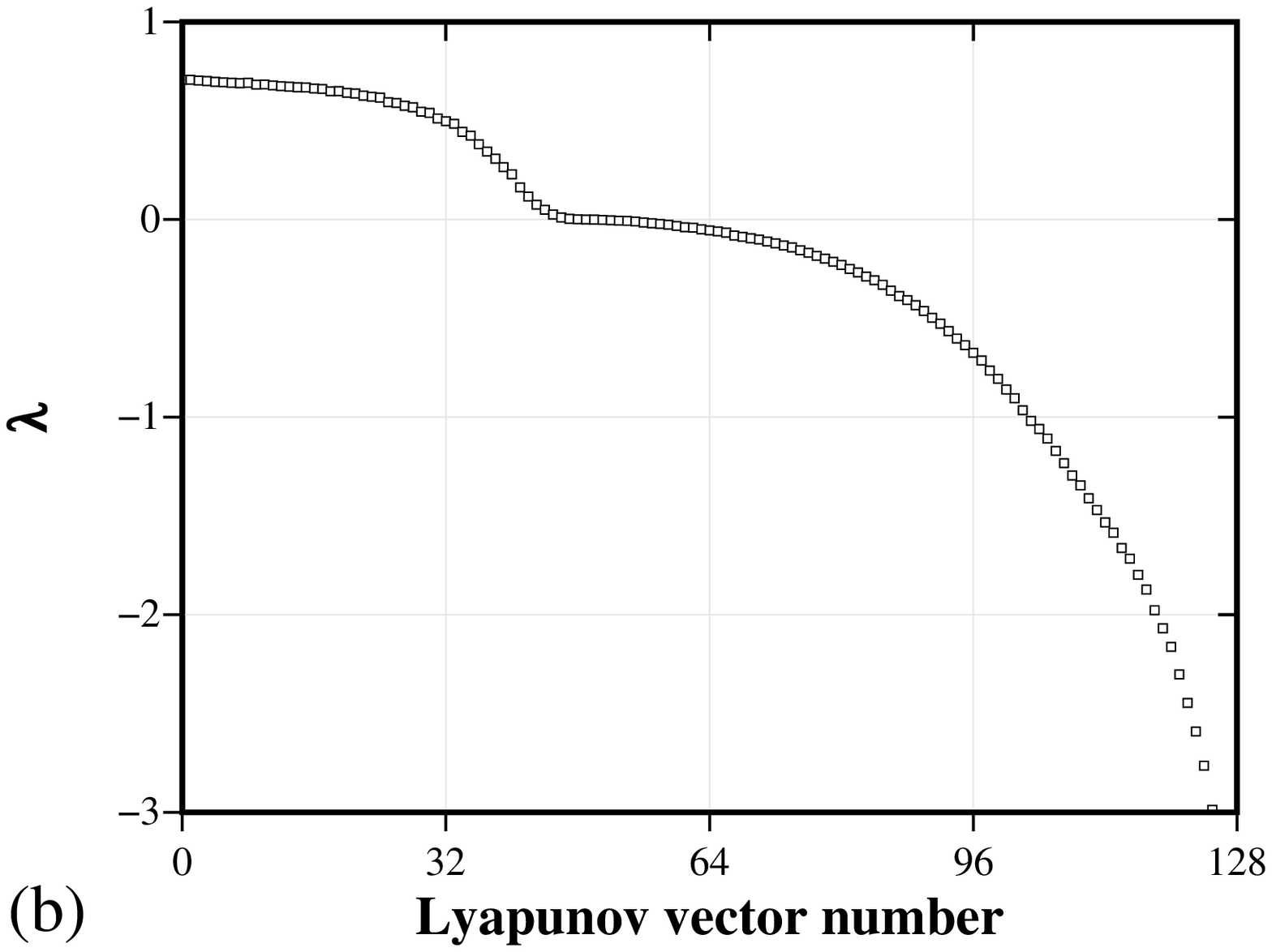,width=3in}}
\vskip 3mm
\caption{
 Power spectrum (a) and Lyapunov spectrum (b) at $u=0.7$, inside the
chaotic phase T1. The dominant contribution to the Lyapunov vectors
corresponding to small $\lambda$ comes from the long-wavelength Fourier
modes only (the white dashed line gives the fit provided by eq. 
(\ref{eq_dominant}) with $\alpha=0.6$). As a result a pronounced
singularity appears in the Lyapunov spectrum (exponent density becomes
singular at $\lambda=0$).  Lattice size is 128.}
 \label{fig_spec_70}
\end{figure}}

\vbox{ 
\begin{figure} 
\centering 
\mbox{
\psfig{figure=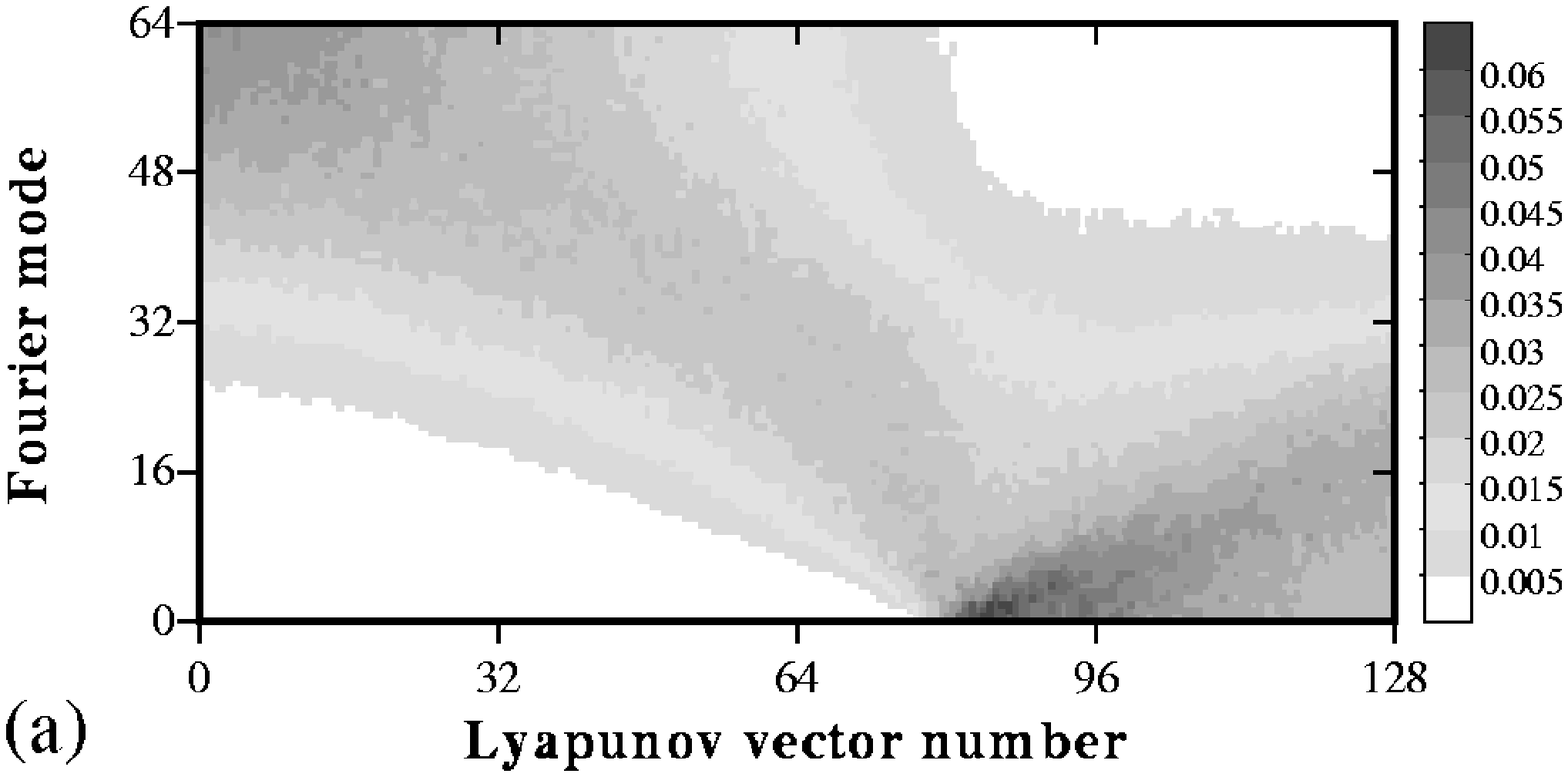,width=3in}}
\vskip 3mm
\centering
\mbox{
\psfig{figure=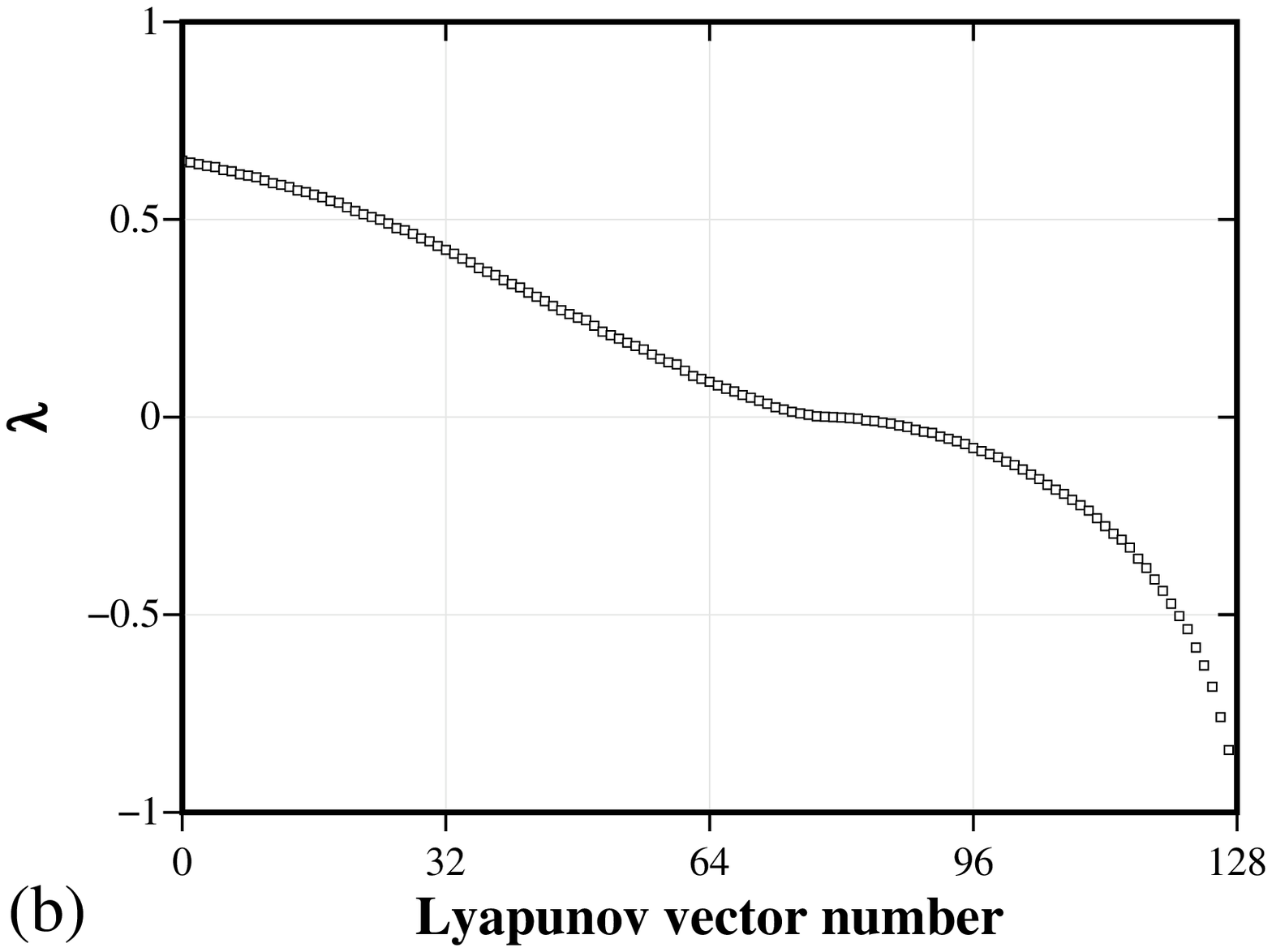,width=3in}}
\vskip 3mm
\caption{
 Power spectrum (a) and Lyapunov spectrum (b) at $u=0.3$, deep inside the
chaotic phase T2. The dominant contribution to the Lyapunov vectors
corresponding to small $\lambda$ comes from the long-wavelength Fourier
modes, but due to the contribution from mid-wavelength modes the
singularity in the exponent density becomes much weaker. Lattice size is
128.}
 \label{fig_spec_30}
\end{figure}}

\vbox{ 
\begin{figure} 
\centering
\mbox{ 
\psfig{figure=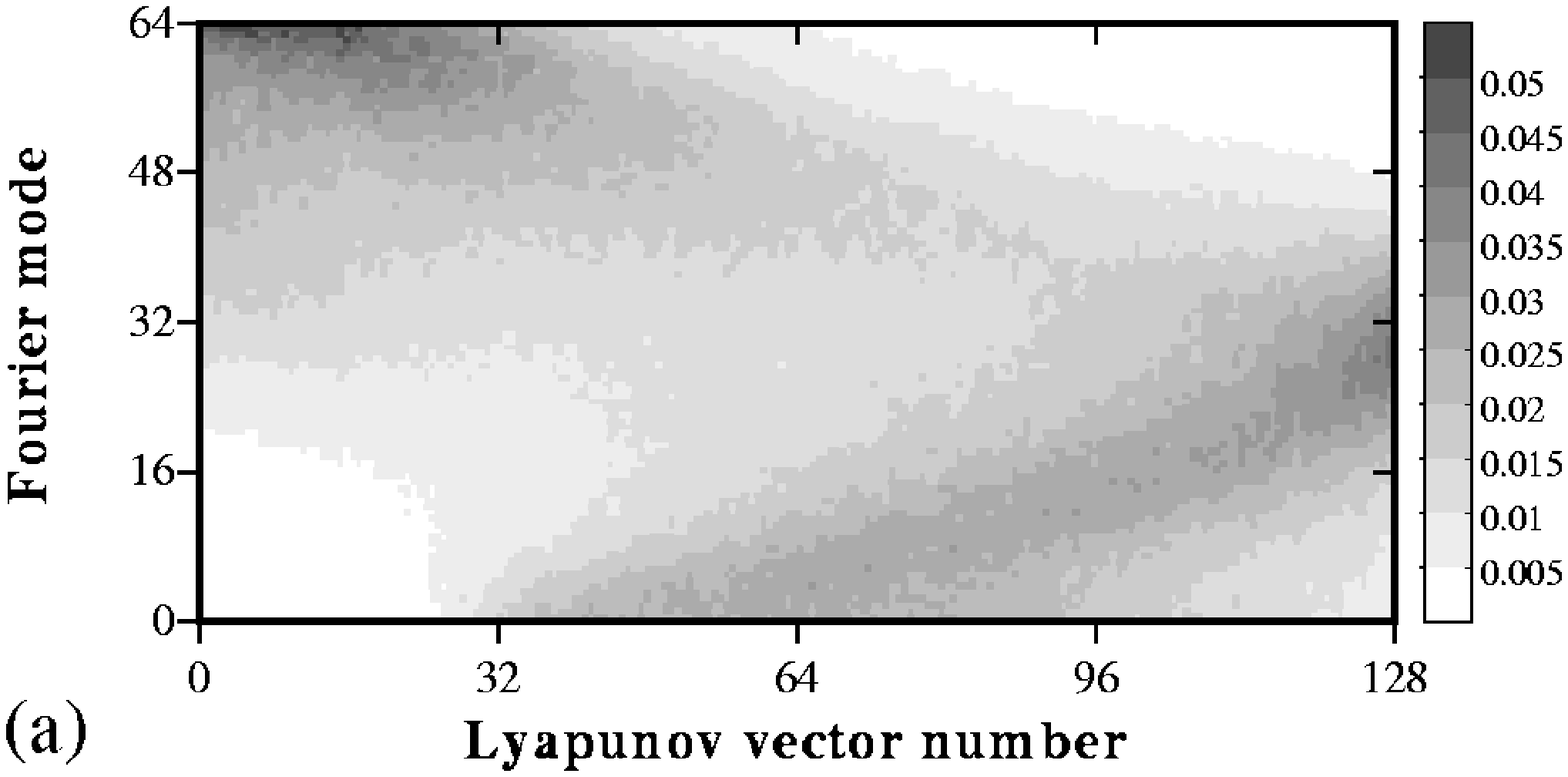,width=3in}}
\vskip 3mm
\centering
\mbox{ 
\psfig{figure=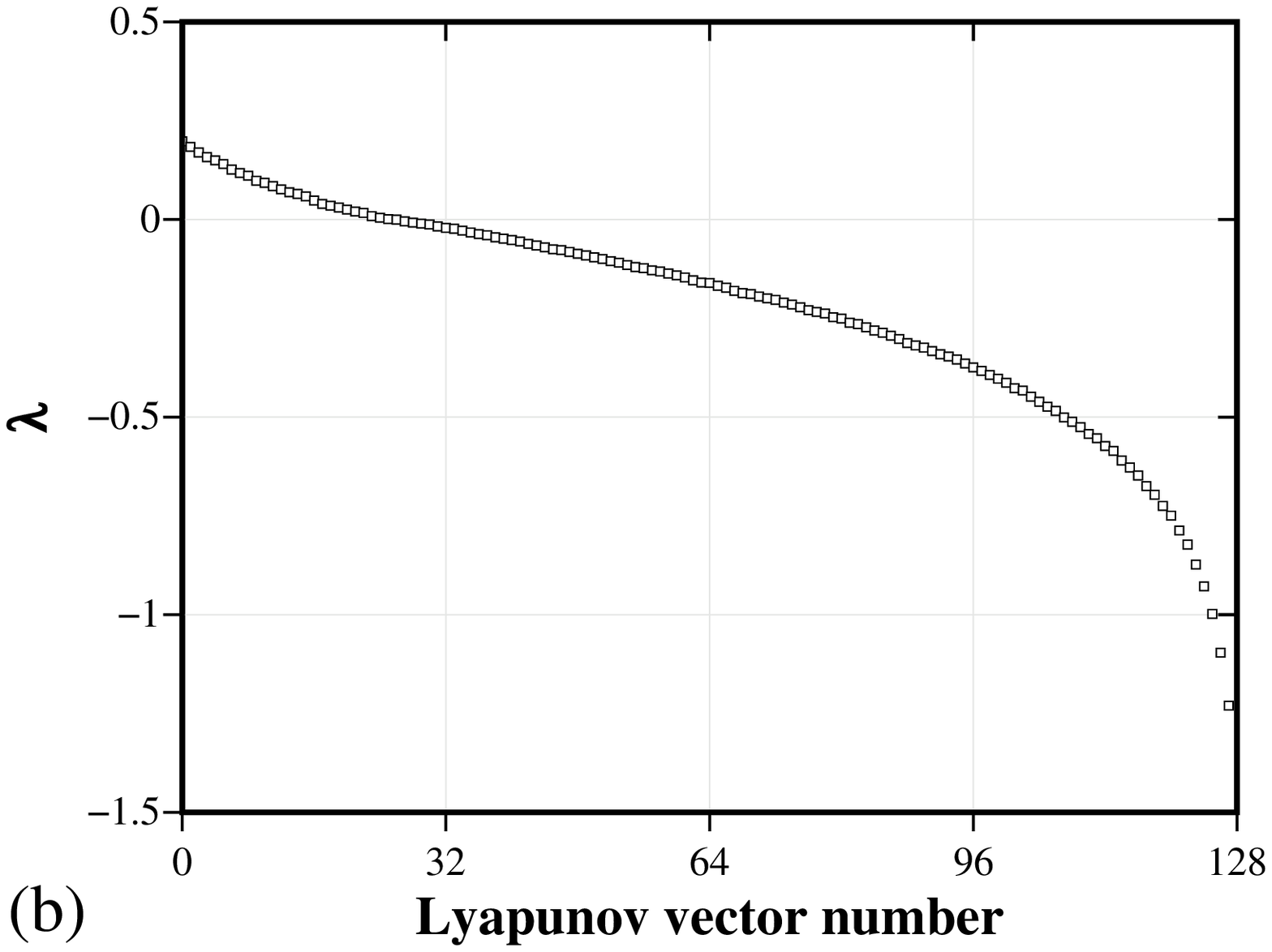,width=3in}}
\vskip 3mm
\caption{
 Power spectrum (a) and Lyapunov spectrum (b) at $u=0.455$, close to the
point where the system experiences the continuous phase transition. There
is a considerable contribution from the short-wavelength Fourier modes to
the Lyapunov vectors corresponding to small $\lambda$ resulting in the
disappearance of the singularity. Lattice size is 128.}
 \label{fig_spec_455}
\end{figure}}

\vbox{
\begin{figure} 
\centering 
\mbox{
\psfig{figure=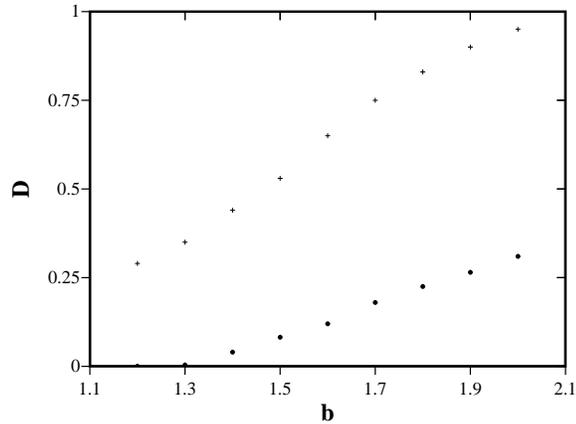,width=3in}} 
\vskip 3mm 
\caption{
 Effective diffusion constant as a function of parameter $b$. $a=0.4$ and
$u=0.8$. The values of $D_{ls}$ ({\scriptsize +}) are determined using the
Lyapunov spectrum of the lattice with $L=128$, assuming $\alpha=0.5$. 
$D_{sf}$ ($\bullet$) is obtained from the dynamic structure function
calculated for $k=0.01\pi$ on the lattice with $L=2048$.}
 \label{fig_diff}
\end{figure}}

\vbox{
\begin{figure}
\centering
\mbox{
\psfig{figure=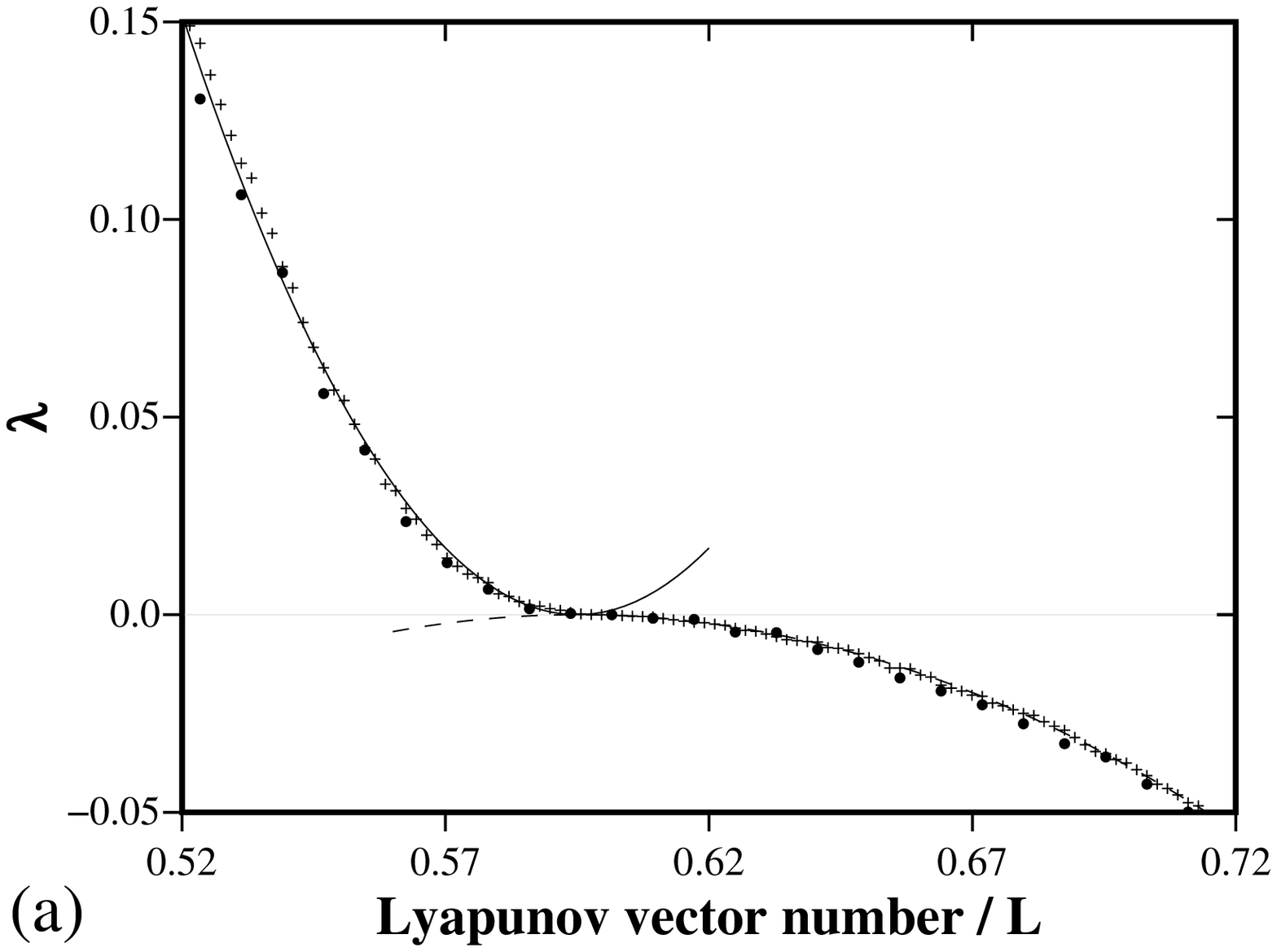,width=3in}\qquad
\psfig{figure=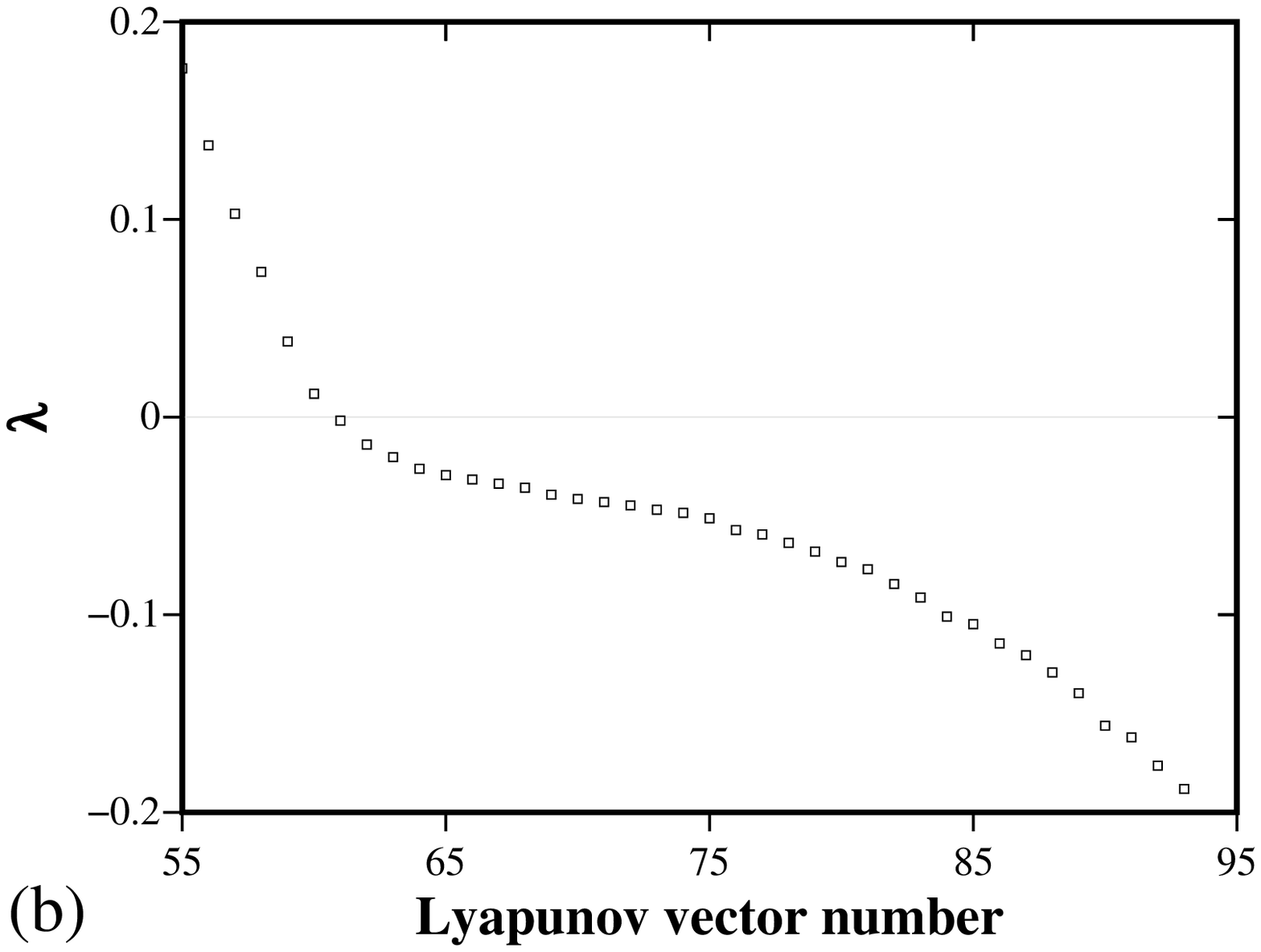,width=3in}}
\vskip 3mm
\caption{
 ``Slow'' part of Lyapunov spectra of the unperturbed system and the
system with nonlinear ``dissipative'' perturbation. (b) The spectrum of
the perturbed system ($u_0=0.8$, $L=128$) with $\epsilon=0.2$ appears to
be shifted downwards and tilted with respect to (a) the spectrum of the
unperturbed system ($u=0.8$). Here the data for two lattice sizes, $L=128$
($\bullet$) and $L=512$ ({\scriptsize +}), is superimposed to show the
finite-size effects. The density of exponents is seen to be singular on
both positive and negative side. The solid and the dashed lines give the
quadratic fit provided by eqs. (\ref{eq_pfit}) and (\ref{eq_spectrum})
respectively.}
 \label{fig_cube}
\end{figure}}

\vbox{
\begin{figure}
\centering
\mbox{
\psfig{figure=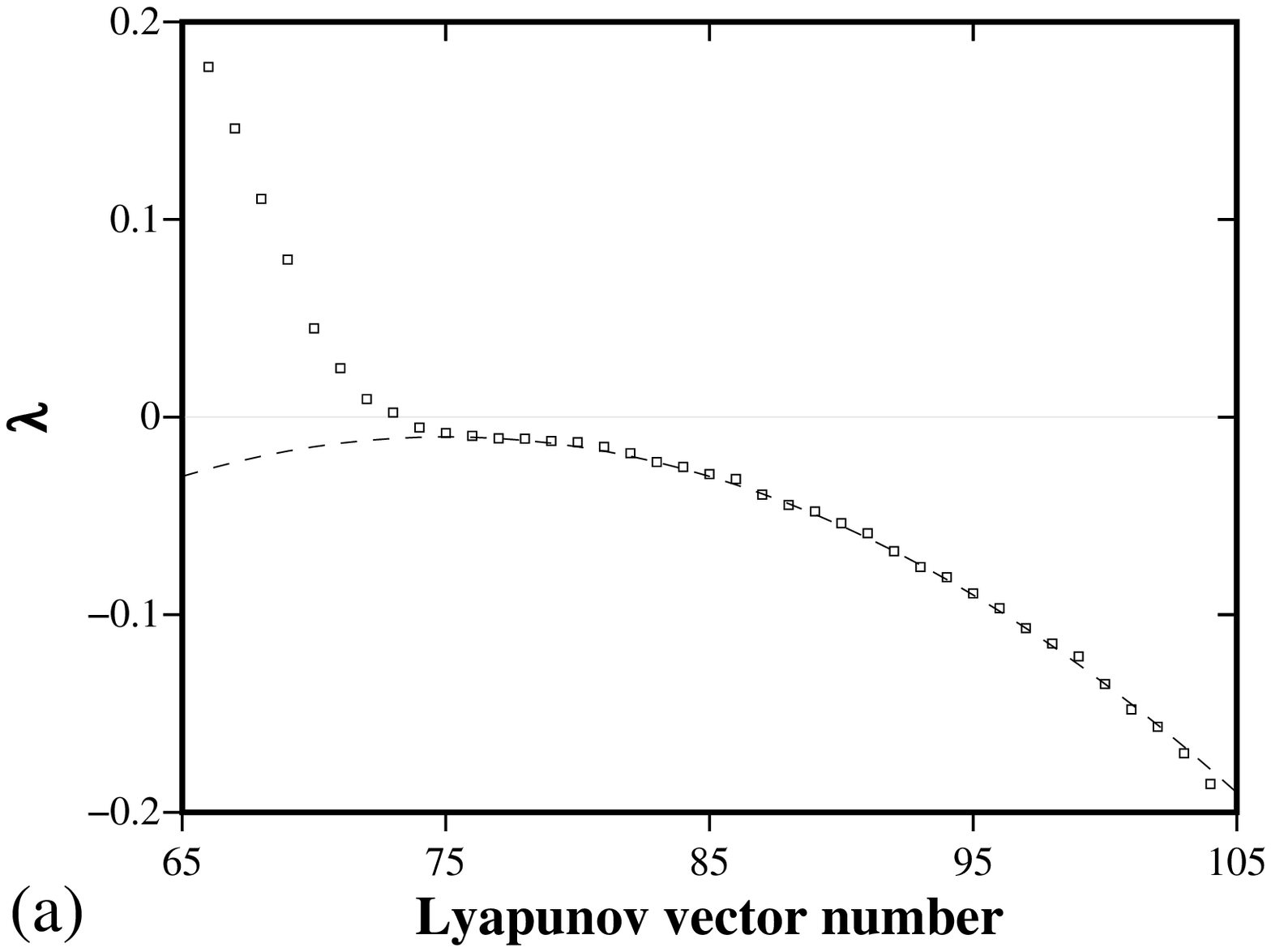,width=3in}\qquad
\psfig{figure=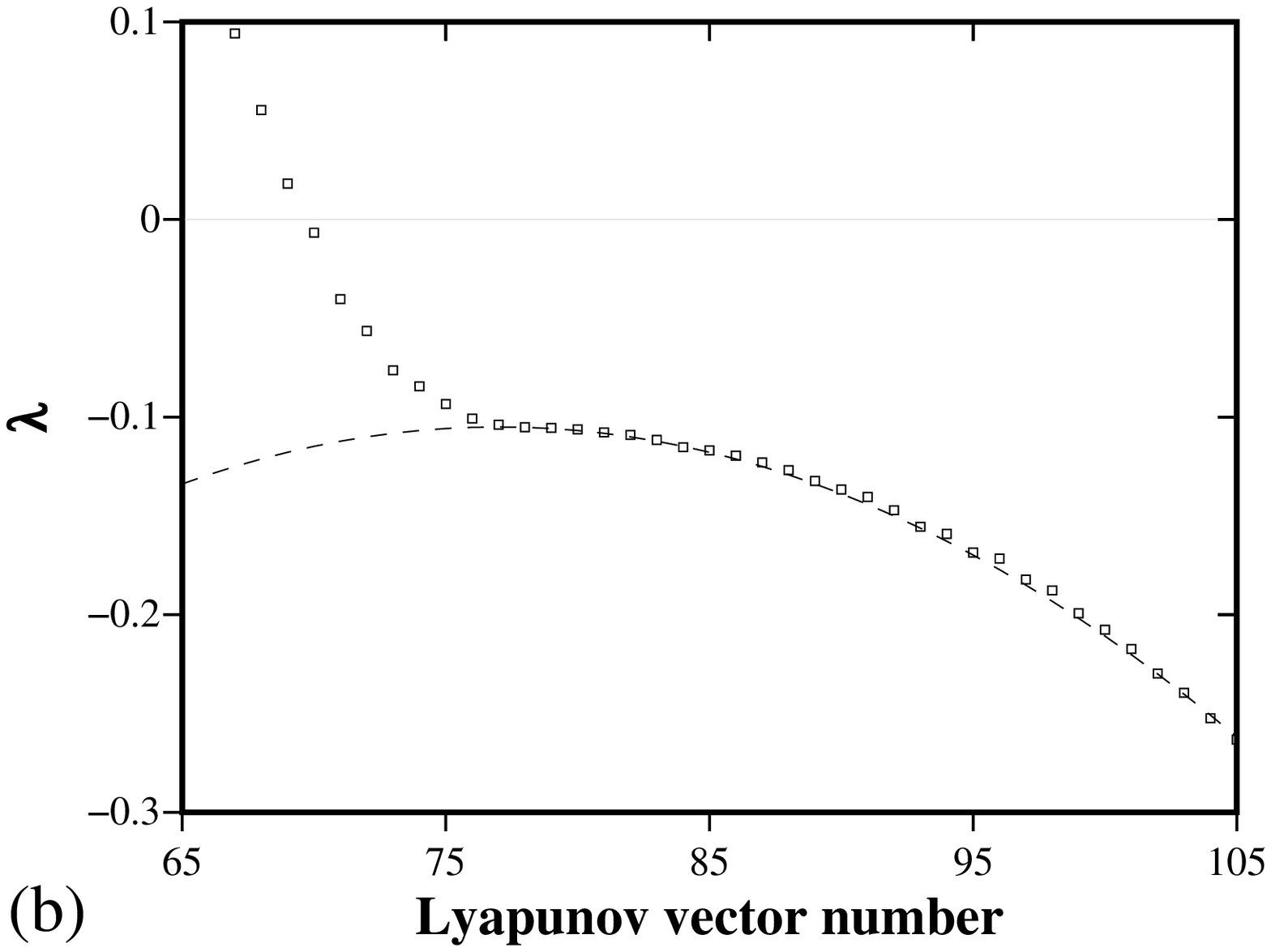,width=3in}}
\vskip 3mm
\caption{
 ``Slow'' part of Lyapunov spectra of the system with linear
``dissipative'' perturbation. For small perturbation (a) $\epsilon=0.01$
as well as for large perturbation (b) $\epsilon=0.1$ the spectrum gets
shifted downwards by $\delta\lambda\approx-\epsilon$ with respect to the
spectrum of the unperturbed system. Lattice size is 128 and $u_0=0.8$. The
dashed line gives the quadratic fit provided by eq. (\ref{eq_qfit}) with
$D=0.33$ (assuming that $k$ is given by eq. (\ref{eq_dominant}) with
$\alpha=0.5$).}
 \label{fig_lin}
\end{figure}}

\vbox{
\begin{figure}
\centering
\mbox{
\psfig{figure=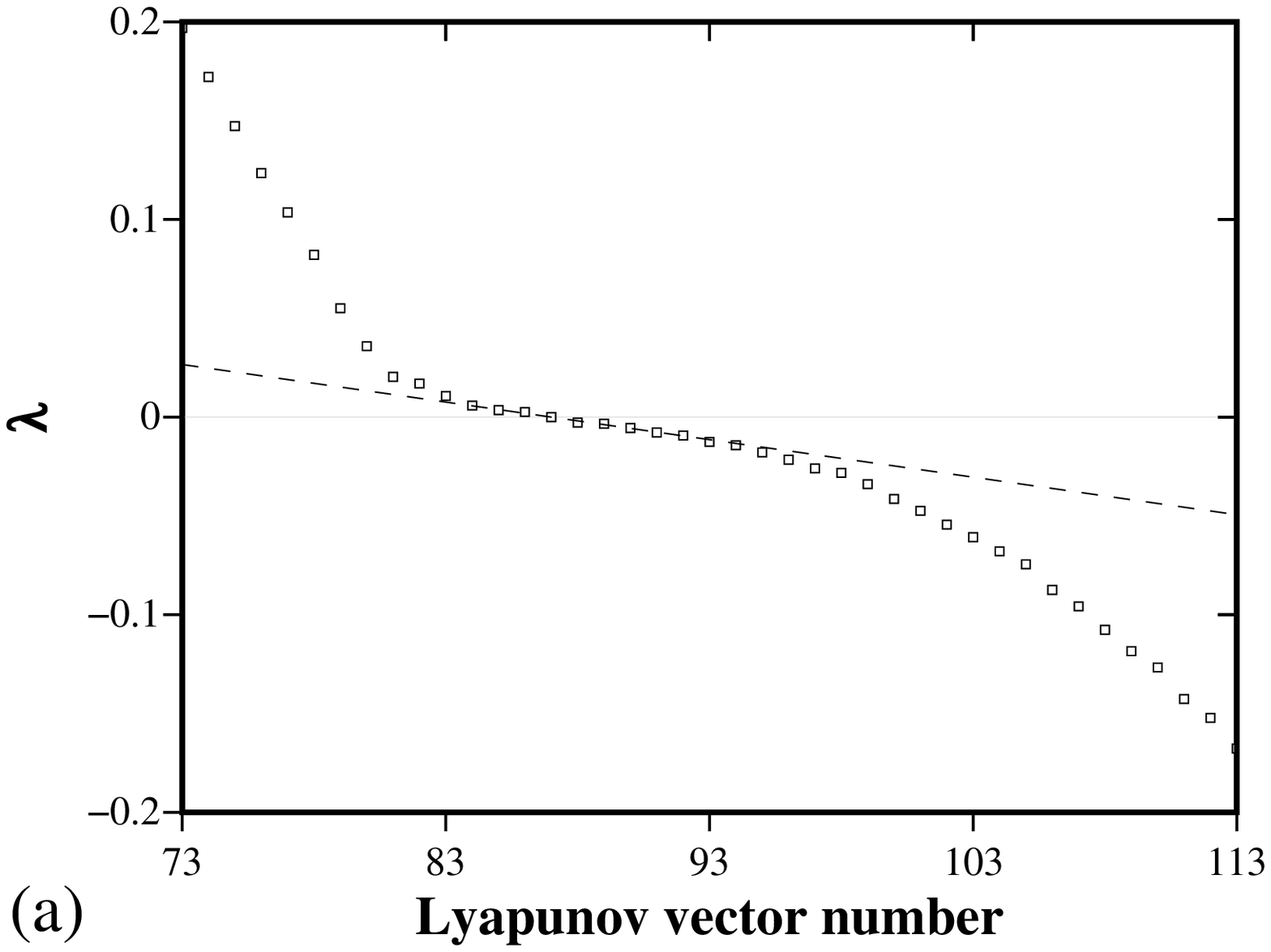,width=3in}\qquad
\psfig{figure=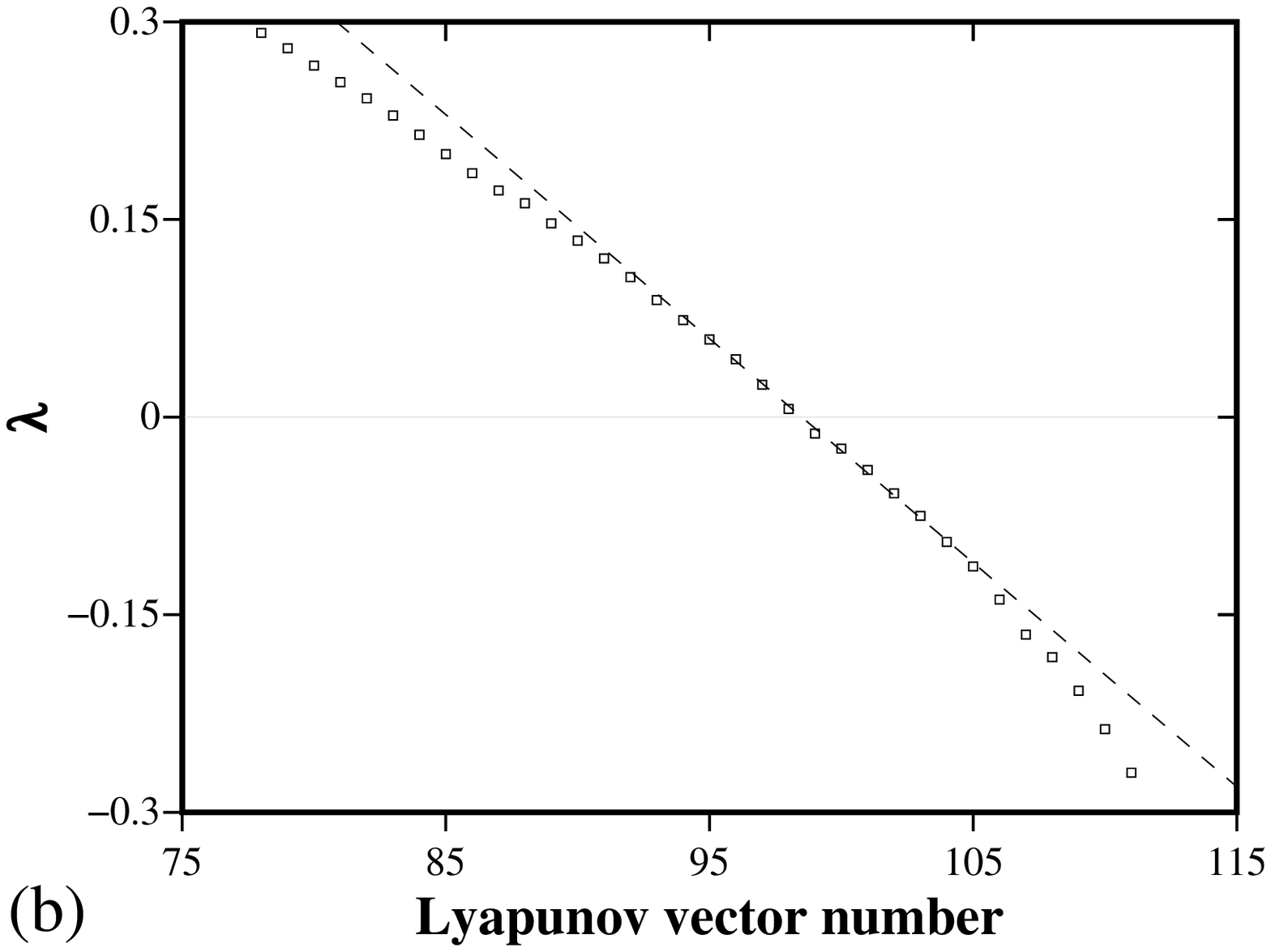,width=3in}}
\vskip 3mm
\caption{
 ``Slow'' part of Lyapunov spectra of the system with ``mixing''
perturbation. For small perturbation (a) $\epsilon=0.1$ the tangent line
(dashed) at the inflection becomes tilted. The singularity disappears. For
very strong perturbation (b) $\epsilon=5.0$ the inflection vanishes
completely. Lattice size is 128 and $u_0=0.8$.}
 \label{fig_lor}
\end{figure}}

\end{document}